\documentclass[aps,twocolumn]{revtex4-1}
\usepackage{graphicx}
\usepackage[justification=justified,width=\linewidth]{caption}
\usepackage{subcaption}
\usepackage{amsmath}
\usepackage{amsfonts}
\usepackage{amsthm}
\usepackage{amssymb}
\usepackage{amsbsy}
\usepackage{wasysym}
\usepackage{bm}
\usepackage{mathrsfs}
\usepackage{color}
\usepackage{times}
\usepackage[resetlabels]{multibib}
\begin{document}
	
	\title{Effect of annealed disorder on the plasticity of amorphous solids}
	\author {Meenakshi L. and Bhaskar Sen Gupta}
	\email{bhaskar.sengupta@vit.ac.in}
	\affiliation{Department of Physics, School of Advanced Sciences, Vellore Institute of Technology, Vellore, Tamil Nadu - 632014, India}
	

\begin{abstract}
We investigate the effect of annealed disorder on the mechanical properties and plasticity of a modeled amorphous solid by introducing a small fraction of heavy impurities into the material which resembles real experimental systems. The disorder being mobile, the total degrees of freedom and the potential energy landscape of the pure system are preserved in our model. The mechanical failure and the shear band formation in the amorphous solid in the presence of annealed disorder are studied at the microscopic level by employing the finite shear deformation protocol at nonzero temperature. A significant enhancement in the shear modulus and yield stress is observed as a function of the heaviness of the impurity particles. Via the analysis of the non-affine displacement field and the microscopic strain fluctuations and the nature of their spatial correlations we find that the shear band formation in the plastic regime is gradually suppressed with increasing impurity mass. Eventually, a critical mass of the disordered particles is identified above which the plastic events become completely localized. This is marked by a transition from a power law to an exponential decay in the spatial correlations non-affine displacement field. Likewise, a similar change is observed in the strain correlation function, transitioning from a slower $1/r$ decay to a more rapid $1/r^3$ decay. Finally, the effect shear rate on the plastic events in the presence of annealed disorder is explored. 
\end{abstract}

\maketitle

\section{Introduction}
The mechanical properties of amorphous materials in the presence of disorder are of paramount importance across numerous scientific and engineering disciplines \cite{chathoth,Demetriou}. Understanding how the disorder affects their mechanical response is crucial for tailoring their properties to meet specific performance requirements \cite{wang}. Impurities in amorphous materials can originate from multiple sources, including fabrication processes, environmental exposure, and intentional doping for specific functionalities. These impurities introduce local structural perturbations, alter the bonding environment, and influence atomic mobility, thereby influencing the material's mechanical response in complex ways \cite{Demetriou1,garett}.

Advancements in experimental techniques and computational modeling have significantly contributed to elucidating the role of impurities in shaping the mechanical properties of amorphous materials. Intentional addition of small amount of impurities to the amorphous materials, often referred to as micro-alloying is used extensively to enhance the properties of the material \cite{gendelman}. This technique is particularly useful in preparing tailored materials with improved physical and chemical properties which includes mechanical properties, glass forming ability, corrosion resistance, thermal stability, magnetic properties and plasticity \cite{wang,chathoth,Demetriou,BMGExp}. Numerous experimental work have endeavored to enhance the mechanical properties of metallic glasses through the incorporation of minute amounts of foreign metals \cite{harmon,garett,parkExperimentalMetallicGlass}. For example, it was demonstrated that an addition of $1\%-2\%$ of Si and Sn in the Cu-Ti based metallic glasses changed the toughness significantly \cite{garett,parkExperimentalMetallicGlass}. Subsequently, computer simulations were performed to understand the effect of disorder via introducing a small amount of randomly pinned particles to glassy systems. An increase in the shear modulus and the toughness of the materials were observed \cite{pankaj2013,pankaj2013-1}.  Recently, the effect of pinning on the elastoplastic behavior and yielding transition of the amorphous materials was investigated at the microscopic level. While for the unpinned glass, the system exhibits spatially localized plastic events by the collective rearrangement of a cluster of particles up to the yielding transition, often referred to as the shear transformation zone (STZ) \cite{argon,falk,Gendelman}, and the subsequent material failure via the system spanning shear band formation, the post yielding scenario completely changed in the presence of pinning \cite{pinaki2019}. The key observations were the change in the yield strain and the gradual decrease in the spatial extent of shear band with increasing pinning concentration. \\
\indent To the best of our knowledge, all the numerical work aimed at understanding the plasticity of amorphous materials affected by impurities were carried out by considering the foreign particles as quenched disorder (infinite mass), i.e. they take part in affine transformation but excluded from non-affine displacement. The deformation  was restricted to the athermal limit with quasistatic strain protocol \cite{pankaj2013,pankaj2013-1,pinaki2019,pinaki2023}.  This prompted us to undertake the current work where we focus on a model for amorphous solids with disorder which closely resembles the real experimental systems. The impurity in the system is modeled here by introducing a small fraction of foreign particles in a glass forming liquid whose mass is finite and increased by a large constant factor. Note that, the impurities in our system resemble anneal disorder which are ergodic in nature. In soft matter, disorder impurities can rearrange themselves, and, therefore, annealed disorder is more prevalent in such systems compared to immobile impurities. \\
\indent We employ the finite shear rate deformation protocol at nonzero temperature. The major differences between our system and the previously used pinned particle systems are the following. Firstly, the total degrees of freedom of all the particles in our system remains the same as the pure system.  Next, unlike the pinning of particles which modify the potential energy landscape of the pure system, with annealed disorder it remains unaffected. Also, the thermodynamic properties of the system remain unaltered and no averaging over the impurity particles is required. Our model allows us to investigate the impact of impurities across various shear rates at finite temperature.  \\
\indent The main goal of this paper is to investigate the effect of annealed disorder particles on the plastic deformation of the amorphous materials when they undergo mechanical loading at finite shear rate and temperature. The mobility of the impurity particles is varied systematically by increasing their mass.  We find the rigidity of the system  measured in terms of the shear modulus increases with the heaviness of the foreign particles. The yielding point also shifts towards higher strain values. One of the important observations post yielding is that the span of the shear band gradually reduces with increasing the mass at a fixed impurity concentration. Finally beyond a critical mass, the plastic events become completely localized, which are homogeneously distributed in space across the system, resembling the pre yield situation. Also, the critical value of the mass reduces with the impurity concentration. The modified plastic response is investigated by studying the characteristics of the non-affine displacement field and the microscopic strain fluctuations and the nature of their spatial correlations in the steady-state regime. Finally, the strain rate dependence of plasticity is investigated in the presence of annealed disorder. 
\section{Model and simulation details}
To prepare the glassy samples, we consider the well studied Kob Anderson binary Lennard Jones system in the NVT ensemble in three dimension \cite{Kob}. $N=150000$ particles are considered in a cubic box with the number ratio $80:20$ of the two species labeled as $A$ and $B$ type. The interaction potential for a pair of particles has the following form
\begin{equation}
	U_{\alpha\beta}(r) = 4\epsilon_{\alpha\beta}\Big[\Big(\frac{\sigma_{\alpha\beta}}{r}\Big)^{12} - \Big(\frac{\sigma_{\alpha\beta}}{r}\Big)^{6} \Big] 
\end{equation}
where $\alpha, \beta \in  {\rm{A, B}}$. The units of various quantities in our simulation are as follows: lengths are expressed in the unit of $\sigma_{AA}$, energies in the unit of $\epsilon_{AA}$, time in the unit of $(m\sigma_{AA}^2/\epsilon_{AA})^{1/2}$ and temperature in the unit of $\epsilon_{AA}/k_{\rm B}$. Here, $k_{\rm B}$ is the Boltzmann constant which is unity. For simplicity the mass of each partcle is taken as unity. The parameters $\sigma_{\alpha\beta}$ and $\epsilon_{\alpha\beta}$ are chosen as follows: $\sigma_{AA} = 1.0, \sigma_{BB} = 0.88, \sigma_{AB} = 0.8$ and $\epsilon_{AA} = 1.0, \epsilon_{BB} = 0.5, \epsilon_{AB} = 1.5$. With these set of parameters, a binary LJ mixture avoids crystallization under supercooling and forms stable glass. The mode coupling transition temperature $T_g$ for this model is approximately $0.44$ in the reduced unit \cite{Kob}. To improve the computational efficiency, the potential is truncated at $r_{cut}=2.5\sigma_{AA}$. 

Molecular dynamics simulation is used to prepare the glassy samples. All simulations are carried out at density $\rho=1.2$ and the system size is taken to be large with total number of particles $N=150000$ to minimize the finite size effect. Positions and velocities of the particles are updated using the velocity-Verlet integration technique \cite{Verlet} with time step $\Delta t =0.005$. The temperature of the system is kept fixed to the desired value by employing the Berendsen thermostat \cite{Berendsen}. Also, we apply periodic boundary conditions in all directions. \\
\indent We begin our simulation by equilibrating the binary mixture in the liquid state at temperature $T=1.0$. The system is then cooled to the final temperature $T=0.2$ below $T_g$  with a fixed cooling rate of $10^{-5}$ to prepare well relaxed glassy samples. To incorporate the disorder in our system, we choose a small fraction of A-type particles and replace them with marked particles denoted by P. Therefore, the mixture comprises A, B, and P types of particles, all interacting via the LJ potential. Barring the mass $m_p$ corresponding to P particles, all other parameters remain the same as A type particles. Finally, the glassy samples are sheared along the $xz$ plane at a constant shear rate $\dot{\gamma}$. Unless otherwise mentioned the results presented in the next section correspond to $\dot{\gamma}= 10^{-4}$. During the shear process, the temperature is controlled by dissipative particle dynamics thermostat \cite{DPD}. The deformed configurations are saved at chosen strain values $\dot{\gamma}t$. For statistical averaging, the whole process is repeated $50$ times with different initial realizations. This ensures the elimination of thermal fluctuation and minimization of error bar of the measured quantities.
\section{Results}
\subsection{The stress-strain curve and shear modulus}
The mechanical properties of the glassy samples in the presence of annealed disorder are explored by imposing a simple shear loading on them at a finite rate. Two different impurity concentrations $c =$ $5\%$ and $10\%$ are considered. The mobility of the impurity particles is varied by changing their mass $m_P$. The simulations are performed with eight different choice of masses, $m_P = 1, 10^2, 10^3, 10^4, 10^5, 10^6, 10^7$ and $10^8$.  

\begin{figure}[ht]
	\centering
	\includegraphics[width=70mm,height=58.5mm]{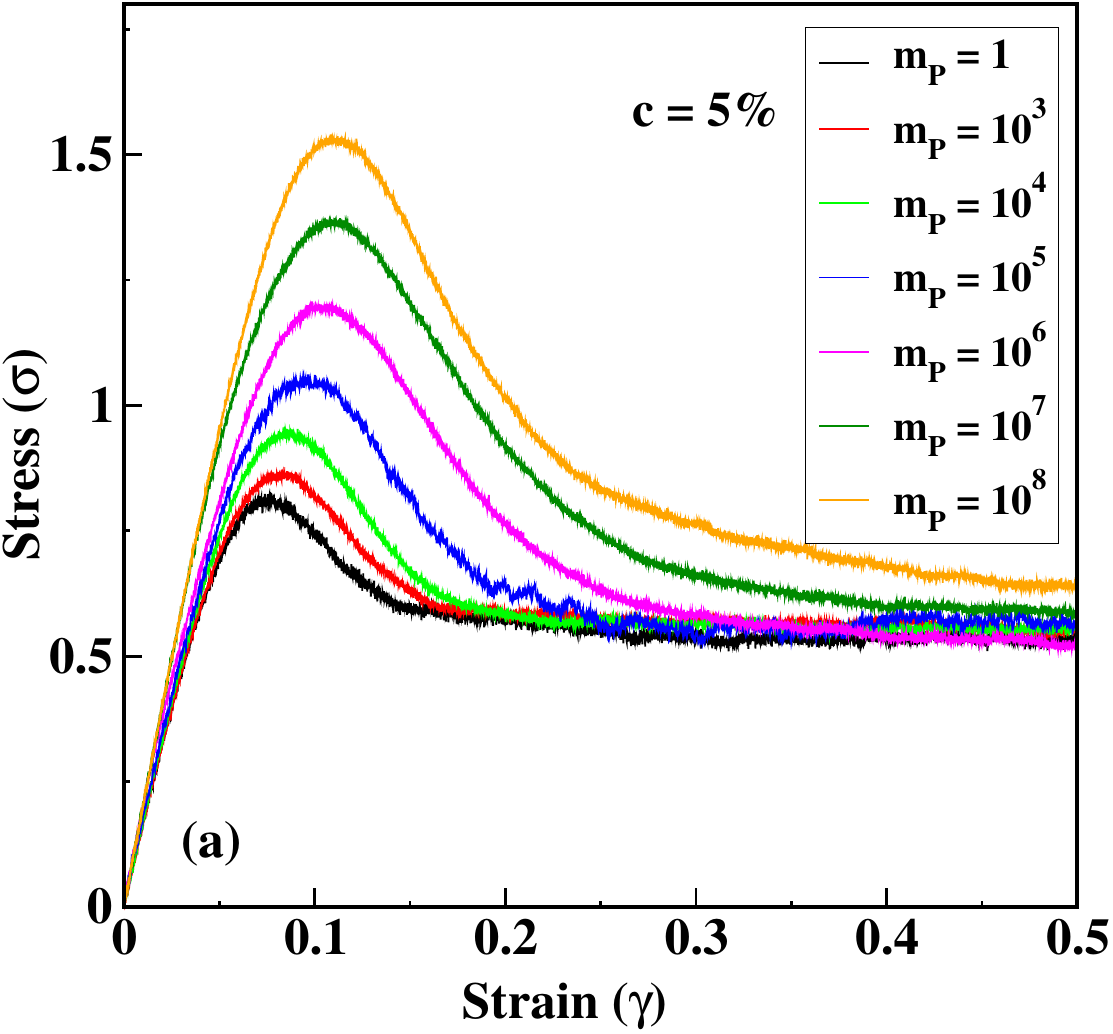}
	\includegraphics[width=71mm,height=58.5mm]{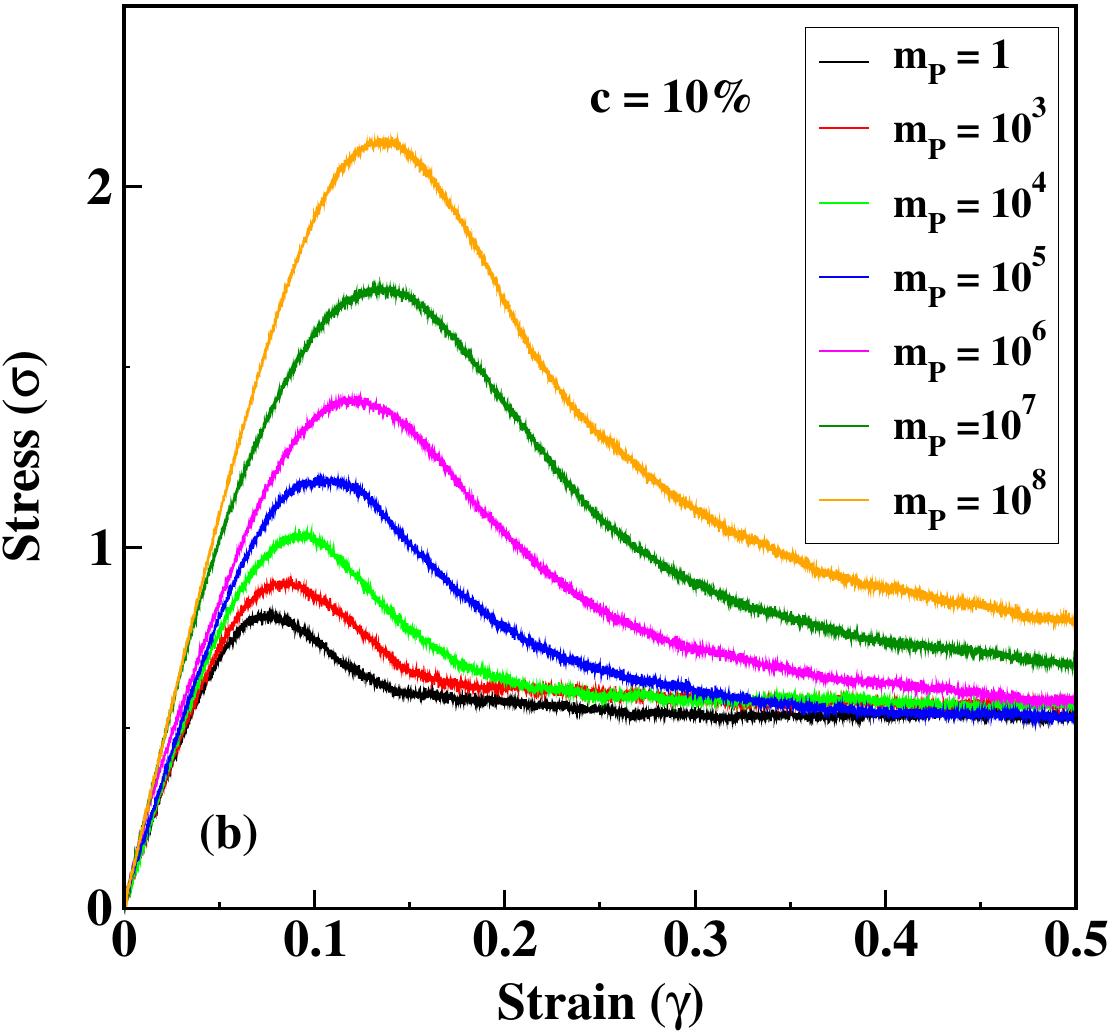}
	\includegraphics[width=70mm,height=58.5mm]{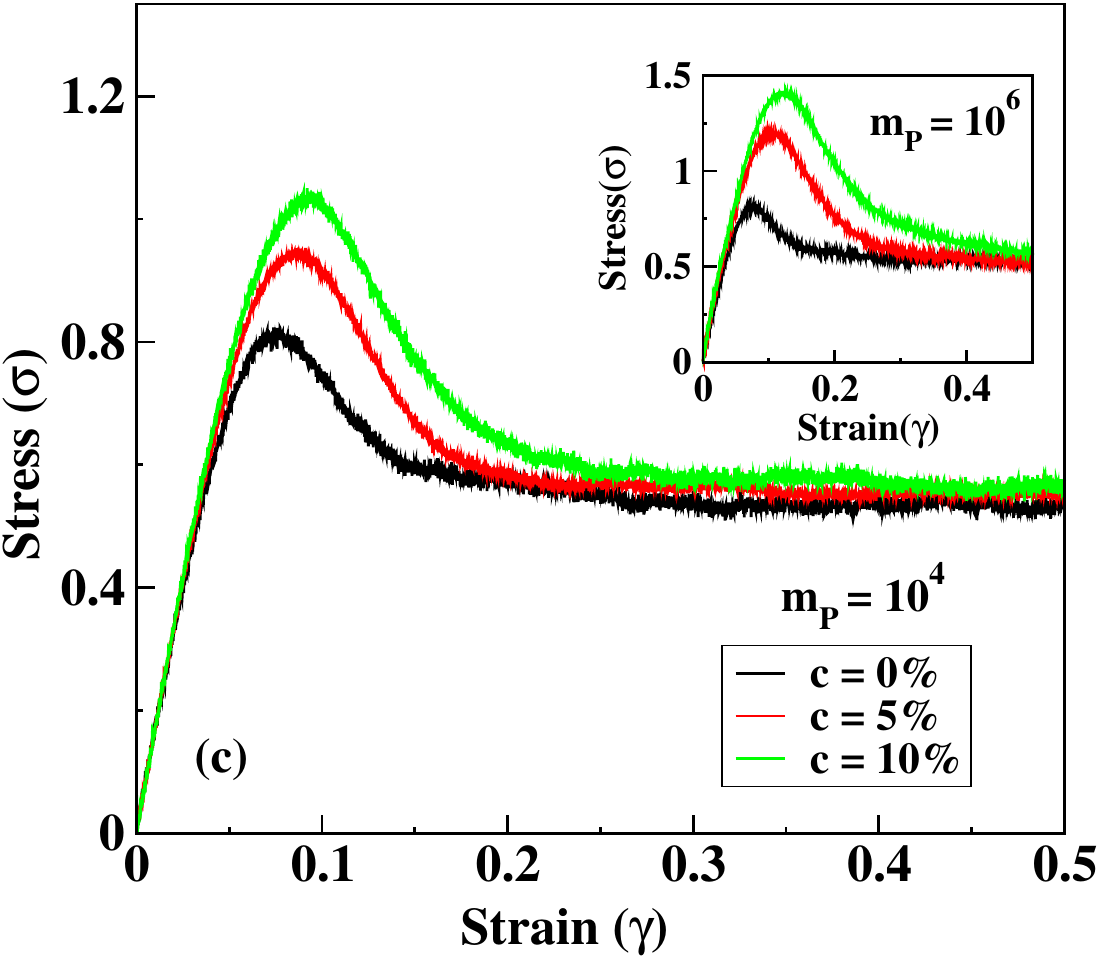}
	\caption{The stress vs strain curves for our glassy system containing (a) 5\% and (b) 10\% annealed disorder particles with different choice of $m_P$ values mentioned in the inset. (c) For comparison, we show the same results for different impurity concentrations at  $m_p = 10^4$ and $10^6$ in the main figure and inset respectively.}
	\label{stressStrain-5percent}
\end{figure}

Figs.~\ref{stressStrain-5percent}(a) and (b) illustrate the evolution of shear stress as a function of shear strain for different $m_P$ values for $c = $ $5\%$ and $10\%$ respectively. In both the cases, for small deformations, stress increases linearly with strain, and then overshoots near yielding before transitioning to the steady-state plastic flow region at higher strain values. From these results, the effect of $m_P$ is conspicuous. With increasing $m_P$, the yielding transition shifts towards higher stress and strain values. The effect of impurity concentration is demonstrated in Fig.~\ref{stressStrain-5percent}(c), which shows stress versus strain curves for two different $c$ values at a specific $m_P$. We find the yielding overshoot becomes more pronounced with impurity concentration, and the location of yielding occurring at larger strain values, similar to the effect of increasing $m_P$. 
\begin{figure}[ht]
	\centering
	\includegraphics[width=65mm]{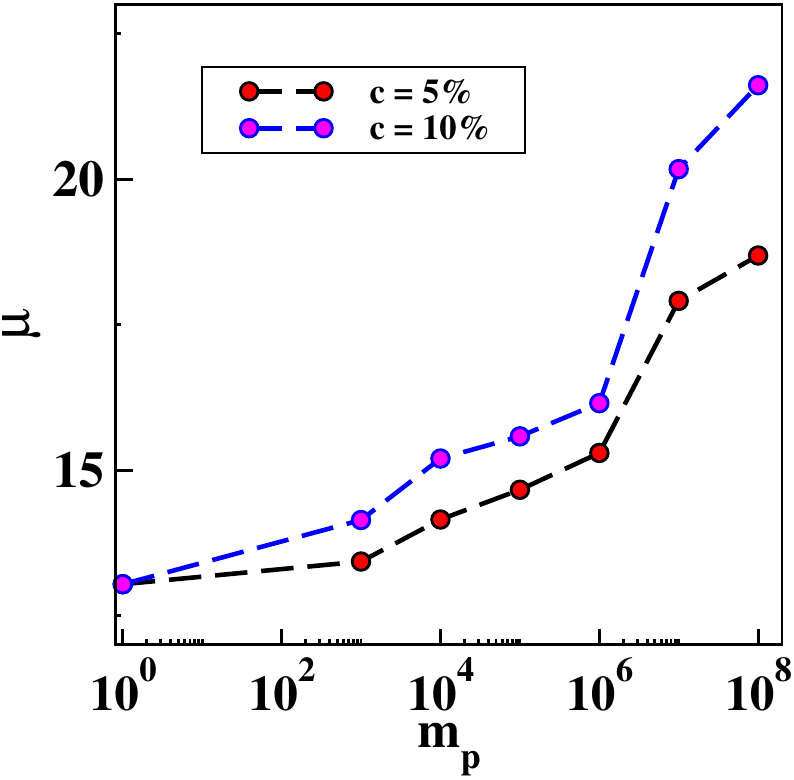}
	\caption{ The variation of shear modulus with the impurity mass $m_p$, for two different disorder concentration.}
	\label{stearMod}
\end{figure}
Meanwhile, under small deformation, the slope of the stress-strain curves becomes steeper with impurity concentrations as well as $m_P$. This is demonstrated in Fig.~\ref{stearMod}. This suggests that the annealed disorder enhances the stability of the system, potentially by augmenting the shear modulus.\\
\subsection{Non-affine displacement field}
\begin{figure}[ht]
	\centering
	\includegraphics[width=40mm,height=35mm]{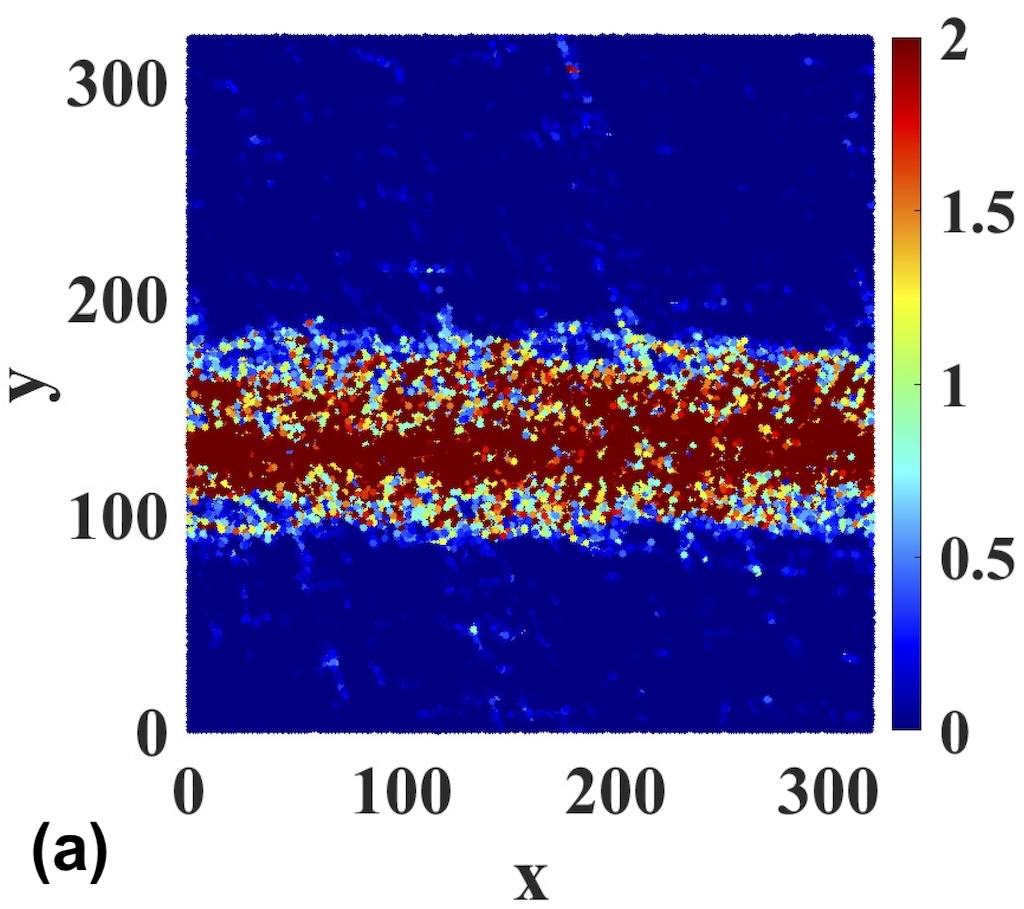}
	\includegraphics[width=40mm,height=35mm]{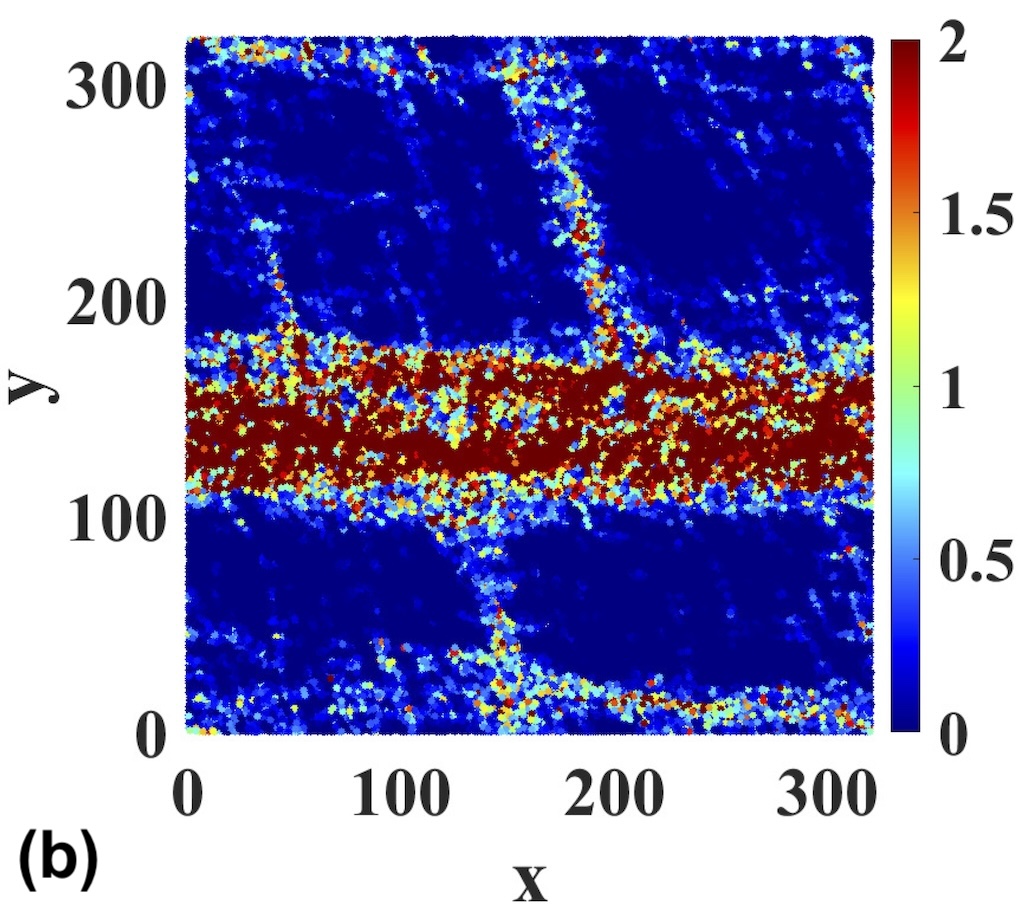}
	\includegraphics[width=40mm,height=35mm]{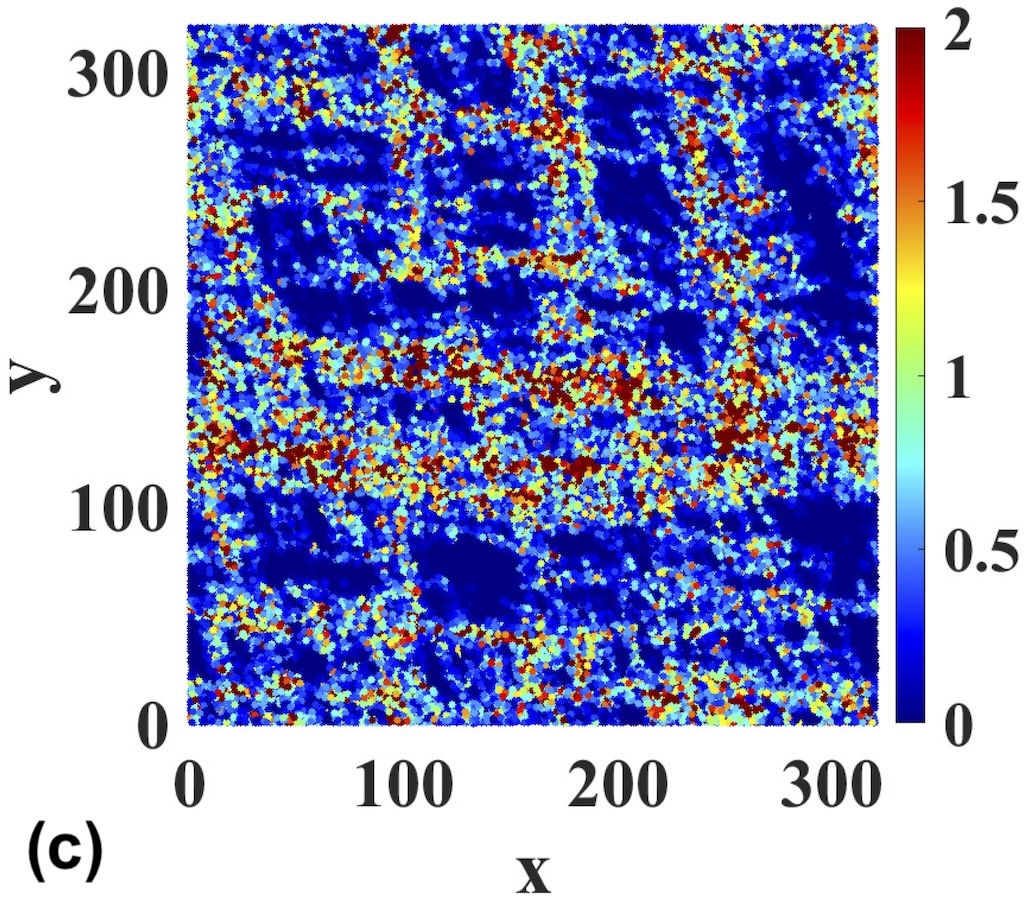}
	\includegraphics[width=40mm,height=35mm]{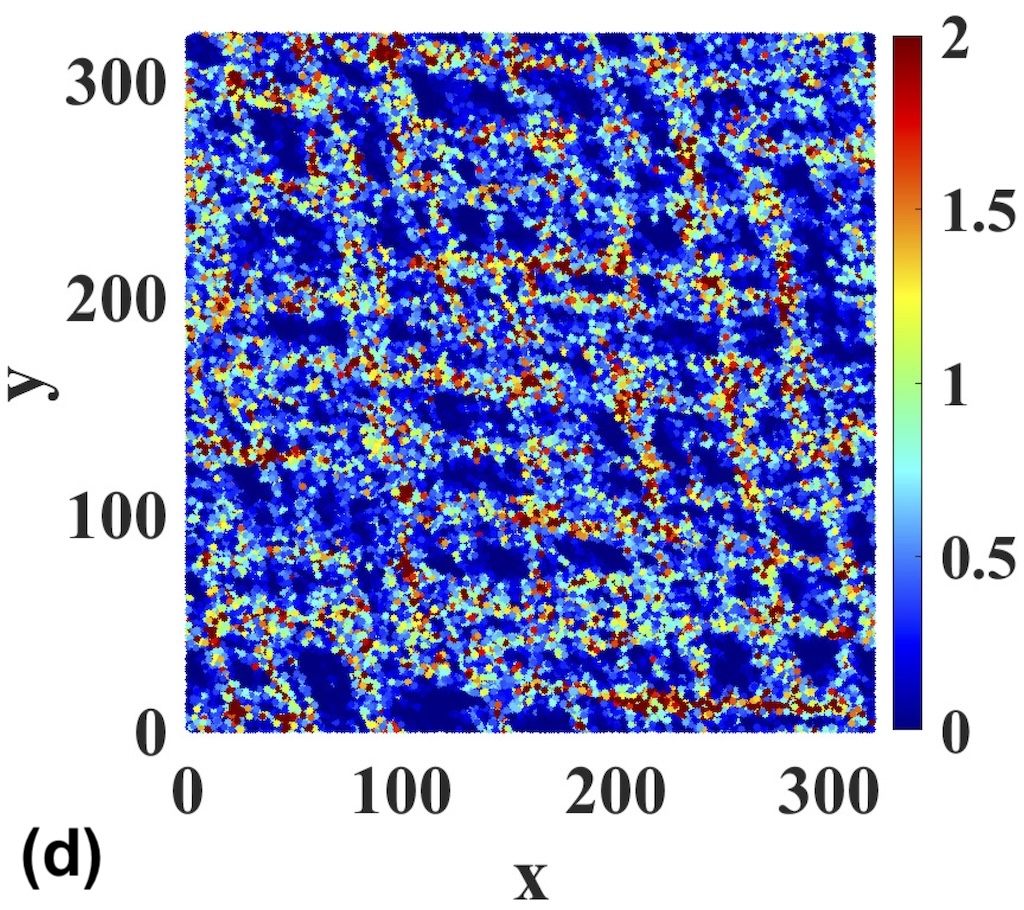}
	\includegraphics[width=40mm,height=35mm]{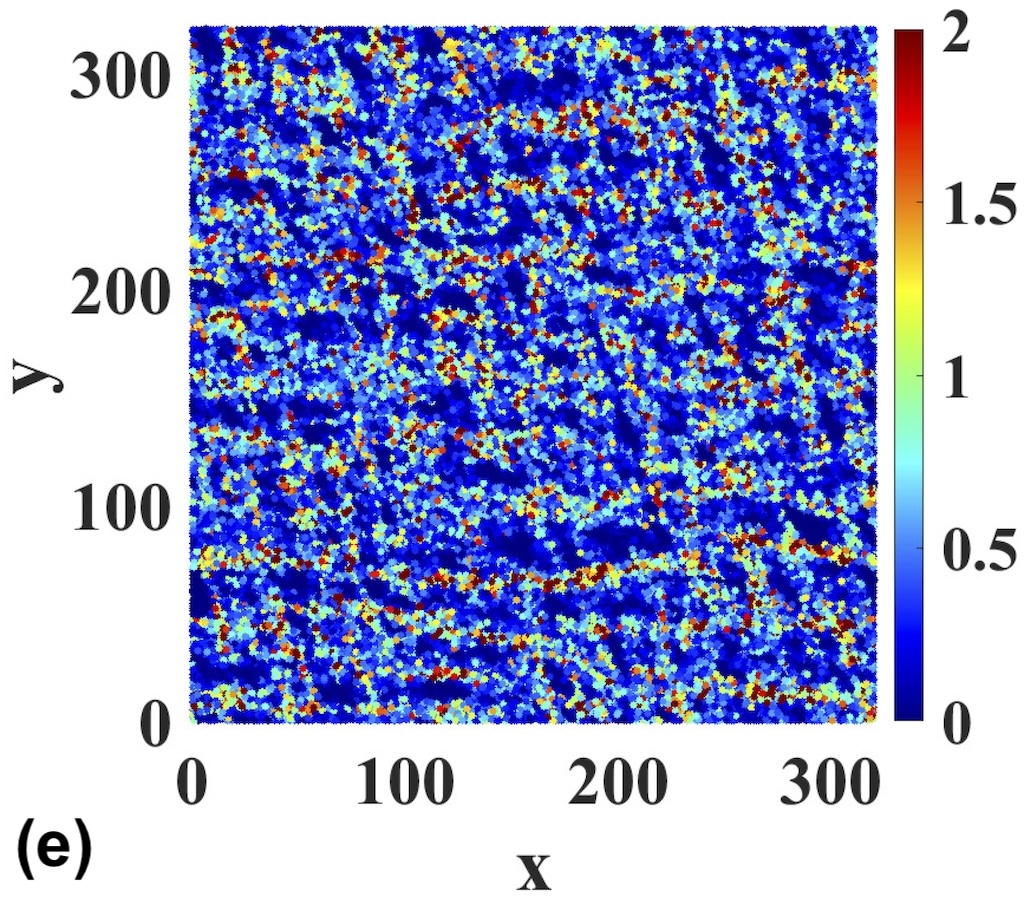}
	\includegraphics[width=40mm,height=35mm]{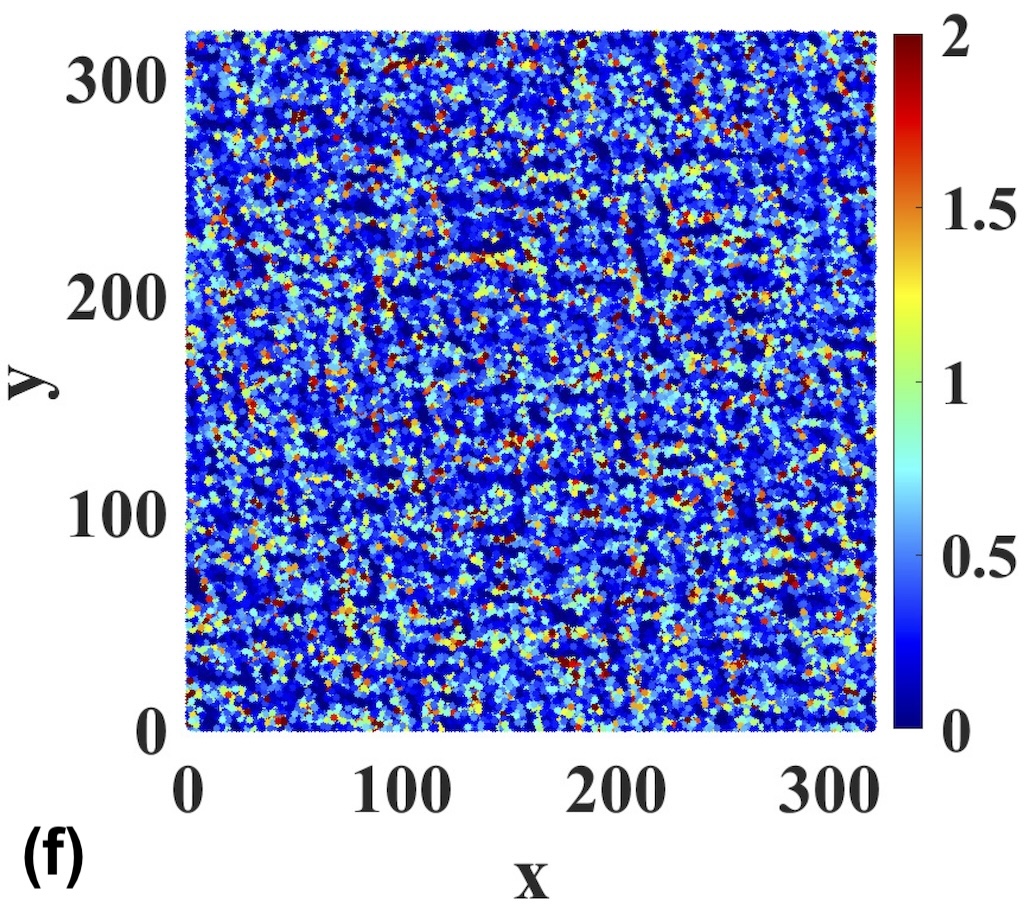}
	\caption{Typical snapshots of the deformed glassy samples for $m_P = 1, 10^3, 10^4, 10^5, 10^6$, and $10^7$ with impurity concentration $c=5\%$ are shown in (a)-(f) respectively. The particles are color-coded according to their non-affine displacements $D^2$ (see text).}
	\label{snap}
\end{figure}

\indent Next we investigate at the microscopic level the influence of annealed disorder on the plastic instabilities occurring in the post yield regime ($\dot{\gamma} t = 0.5$). For that, we resort to the microscopic analysis involving spatial configurations of atoms with large relative non-affine displacements associated with the deformation events. More precisely, we compute the non-affine displacements of the local shear transformations in the glassy samples, originally proposed by Falk and Langer given by \cite{falk}
\begin{equation}
	\begin{split}
D^2(t,\Delta t)  = & \frac{1}{N_i} \sum\limits_{j=1}^{N_i}  \lbrace \boldsymbol{\mathrm{r}}_j(t) - \boldsymbol{\mathrm{r}}_i(t) - \\
&   \boldsymbol{\mathrm{J}}_i [\boldsymbol{\mathrm{r}}_j(t-\Delta t) - \boldsymbol{\mathrm{r}}_i(t-\Delta t)]\rbrace^2
		\end{split}
	\label{eq-d2}
\end{equation}
Here $\boldsymbol{\mathrm{r}}_i(t)$ denotes the position vector of the $i$th particle at the specific strain value $\dot{\gamma} t$ at time $t$ , $\boldsymbol{\mathrm{J}}_i$ is the transformation matrix that maps the $i$th particle and its nearest neighbors at strain $\dot{\gamma}t$ and $\dot{\gamma}(t-\Delta t)$ through an affine deformation.  This quantity is extensively used for the spatio-temporal analysis of the non-affine displacements in shear-driven amorphous solids \cite{jana,chikkadi,varnik,ding,meenakshi,chikkadi2011,nikolai2016,nikolai2017,santhosh,niyogi}. The time window considered here corresponds to $\dot{\gamma}\Delta t = 0.01$. Because of the choice of small window, the results presented here are insensitive to the particular time interval.  \\
\indent We begin the analysis by showing the representative snapshots for the two dimensional deformed glassy systems containing $5\%$ annealed disorder in the steady state regime in Fig.~\ref{snap} for the chosen $m_P$ values. Reduced dimensionality is opted here for better visualization. The two dimensional glassy systems involve $N=100000$ particles with 65:35 concentration ratio for A and B types. The interaction parameters and preparation protocol remain consistent with those utilized for the three-dimensional glassy samples. The corresponding glass transition temperature in this case is $T_g = 0.44$ \cite{2dGlass}.  Next, we compute the displacement field $D^2 $ for these configurations from Eq.~\ref{eq-d2}. In Fig.~\ref{snap} we color-code the particles according to their non-affine displacement field. 
For the pure system, accumulated plastic events drive the formation of system-spanning shear band exhibiting avalanche behavior. This process leads to material failure and the heterogeneous distribution of the non-affine displacement field, as indicated by the color-code. The presence of annealed disorder in the system induces a transformation in the extend and spatial distribution of the plastic activities. Increasing the mass $m_P$ gradually shrinks the spatial scale of regions experiencing significant non-affine displacements. Finally, above a certain critical mass $m_c$, the shear band is completely suppressed and the plastic events become localized, exhibits a more homogeneous distribution throughout the system.\\
\indent Our observations reveal the critical mass $m_c = 10^5$ for the shear rate  $\dot{\gamma}=10^{-4}$, a coexistence of delocalized shear band formation and localized plastic activities are observed as shown in the Fig. \ref{snap}. Notably, shear bands are suppressed above this $m_c$. Further increment of $m_P$ induces continued pervasion of localized plastic rearrangements across the entire sample, resulting in a homogeneously distributed plastic activities.\\

\begin{figure}[ht]
	\centering
	\includegraphics[width=40mm,height=35mm]{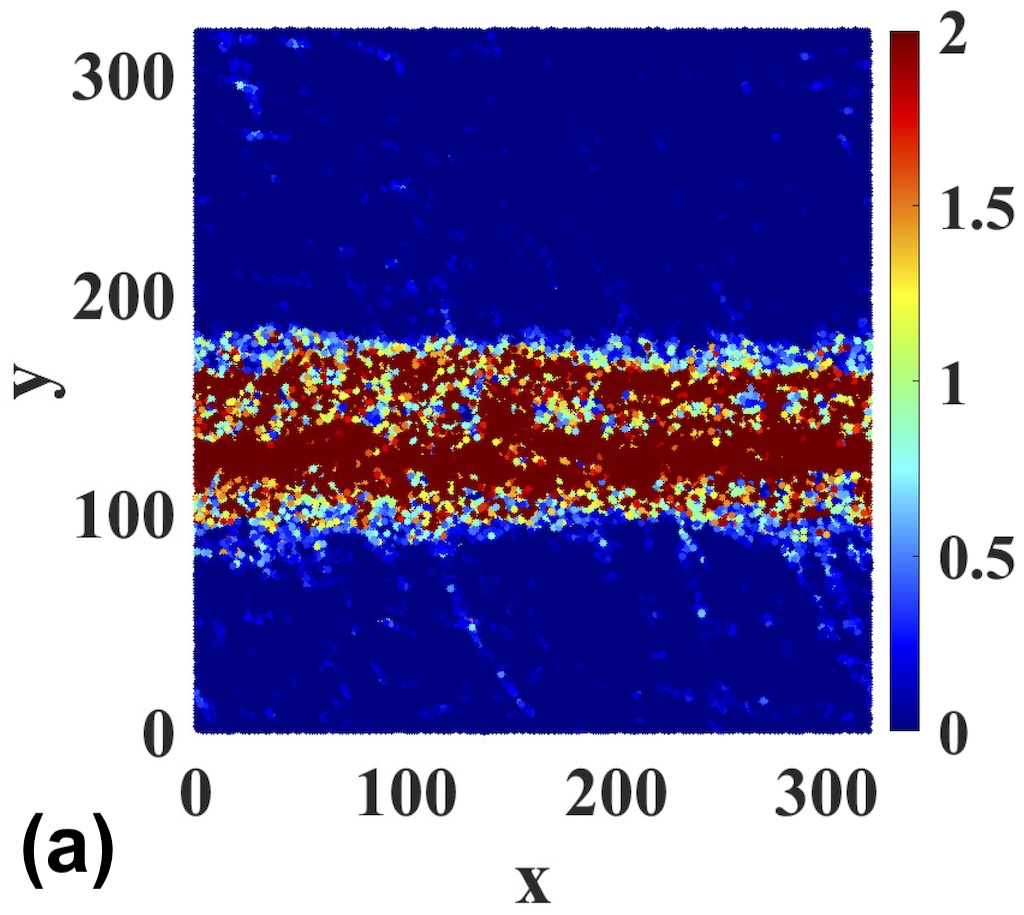}
	\includegraphics[width=40mm,height=35mm]{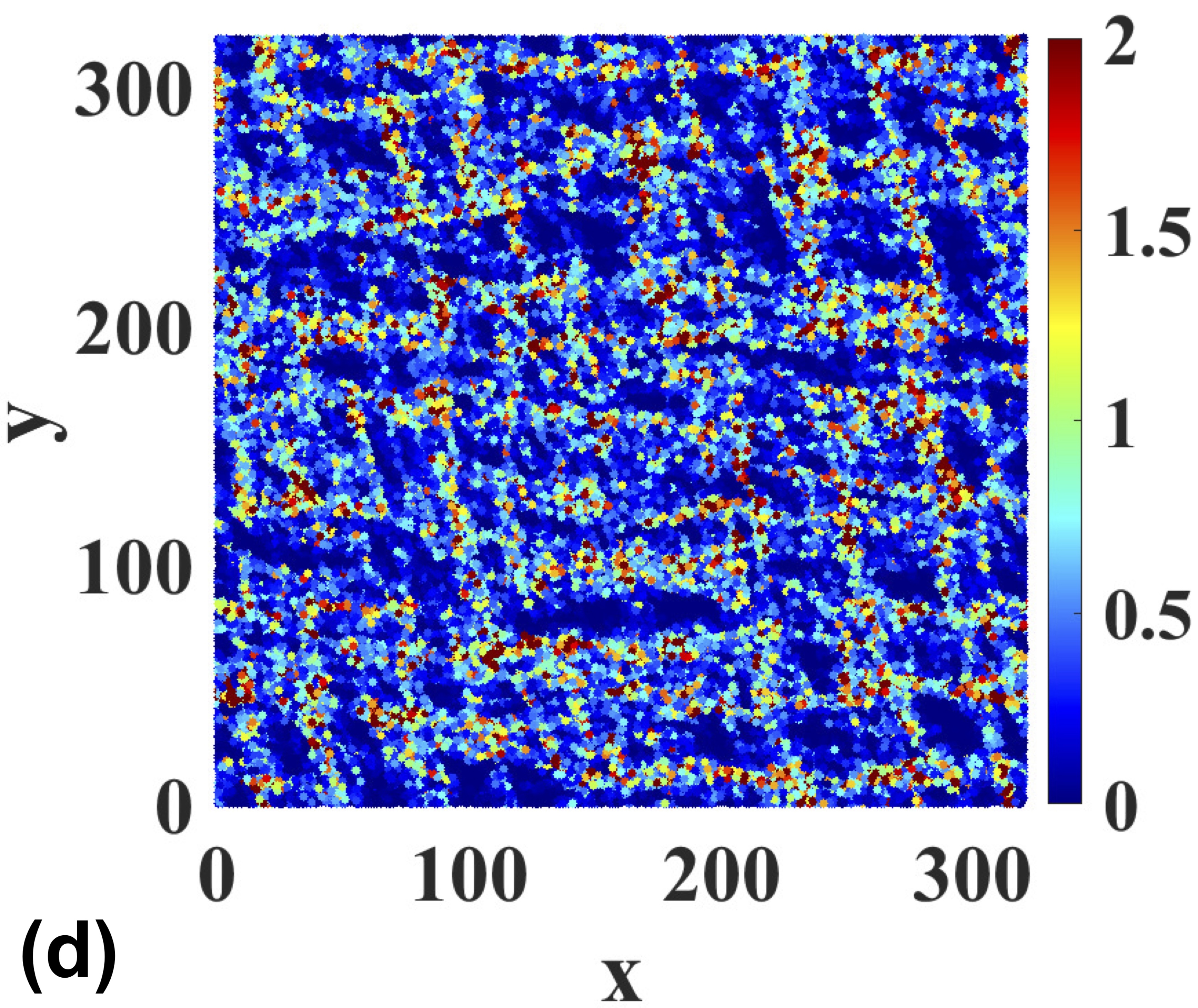}
	\includegraphics[width=40mm,height=35mm]{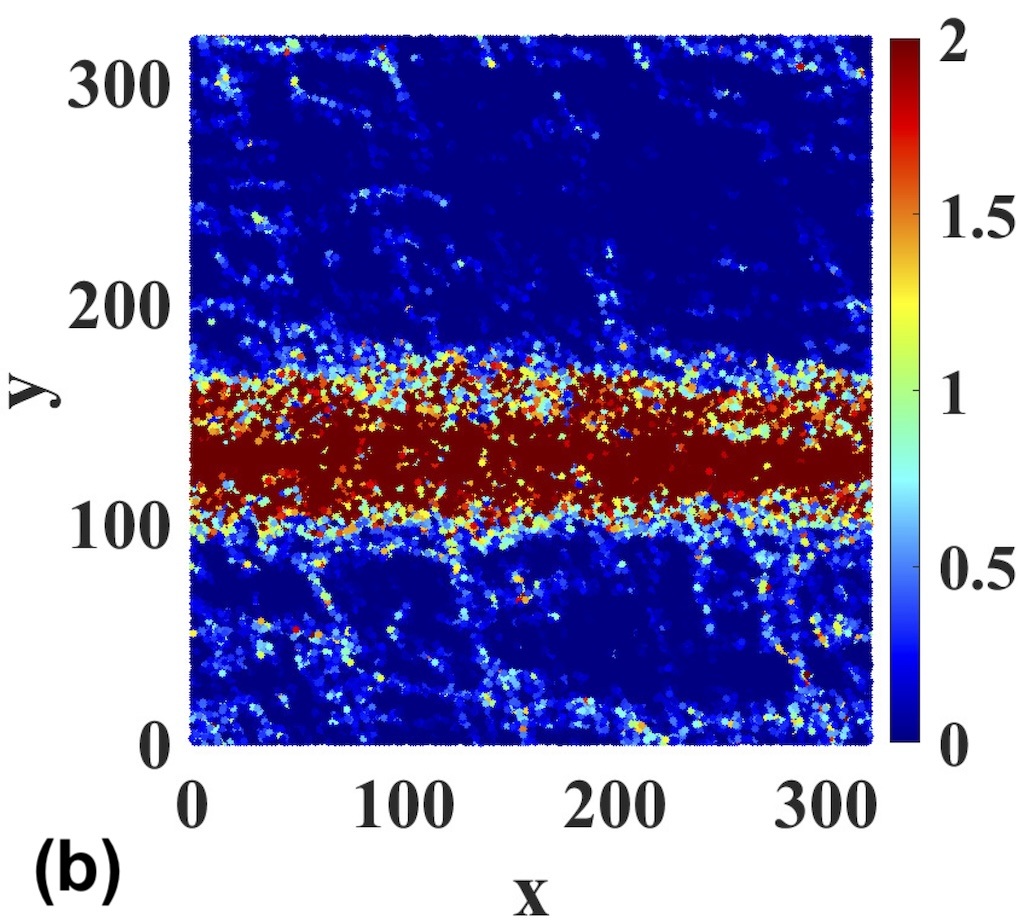}
	\includegraphics[width=40mm,height=35mm]{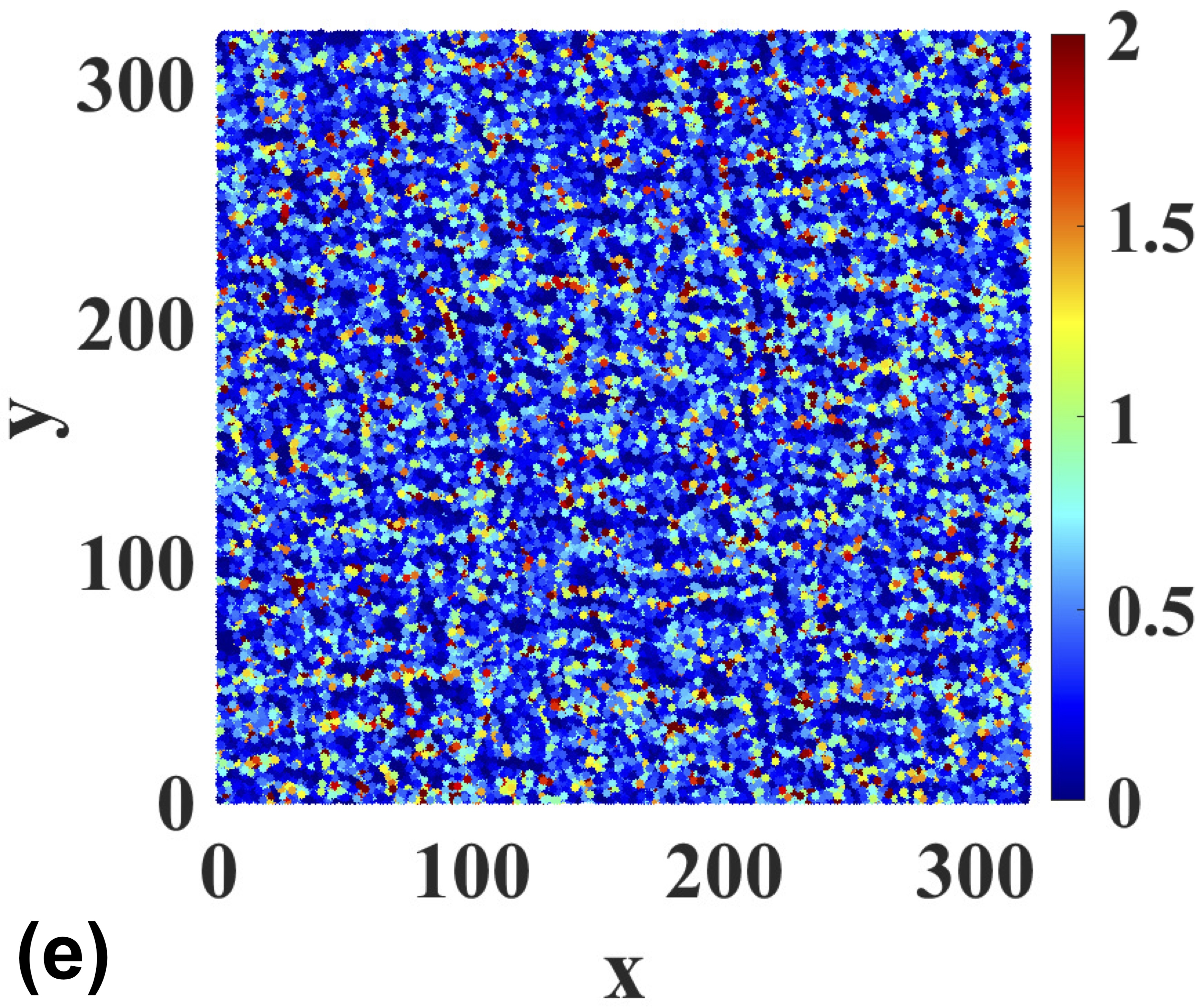}
	\includegraphics[width=40mm,height=35mm]{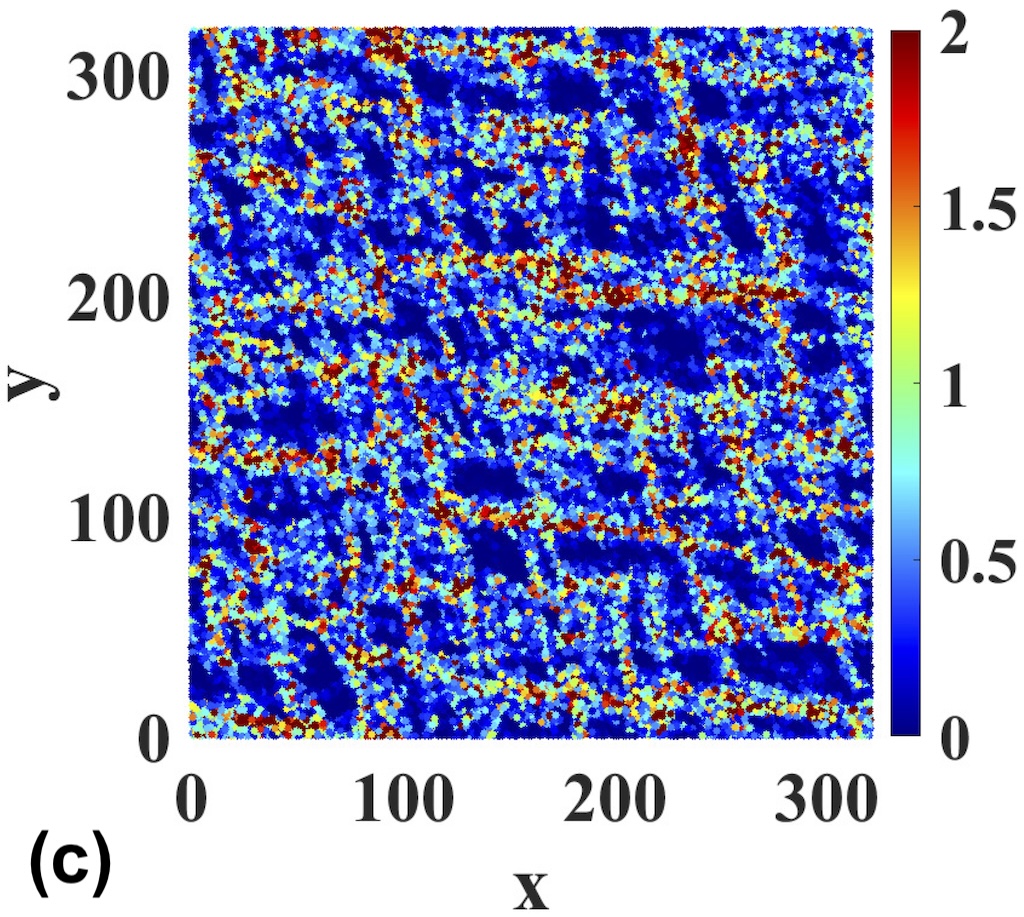}
	\includegraphics[width=40mm,height=35mm]{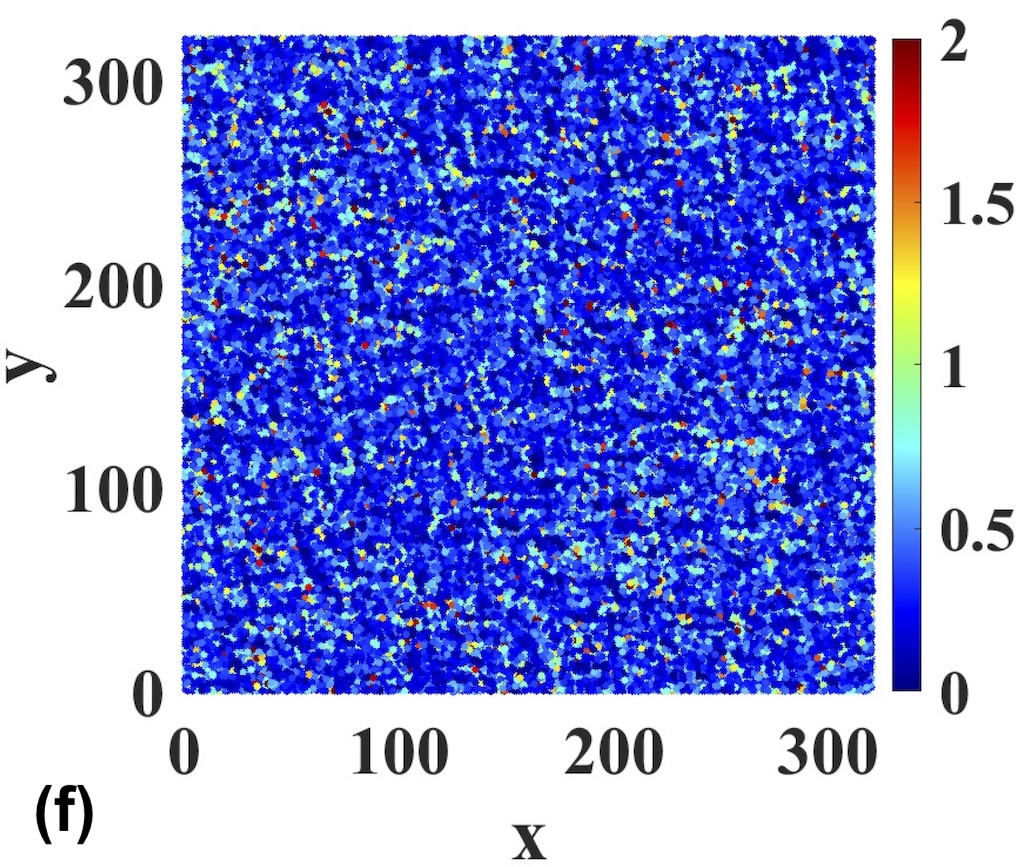}
	\caption{Typical snapshots of the deformed glassy samples for $m_P = 10^3, 10^5$ and  $10^7$,  with impurity concentration $c=5\%$ are shown in (a)-(c) for $\dot{\gamma}=10^{-5}$ and (d)-(f) for $\dot{\gamma}=10^{-3}$ respectively. The particles are color-coded according to their non-affine displacements $D^2$ (see text).}
	\label{snaperate}
\end{figure}

\begin{figure}[hbt]
	\centering
	\includegraphics[width=60mm]{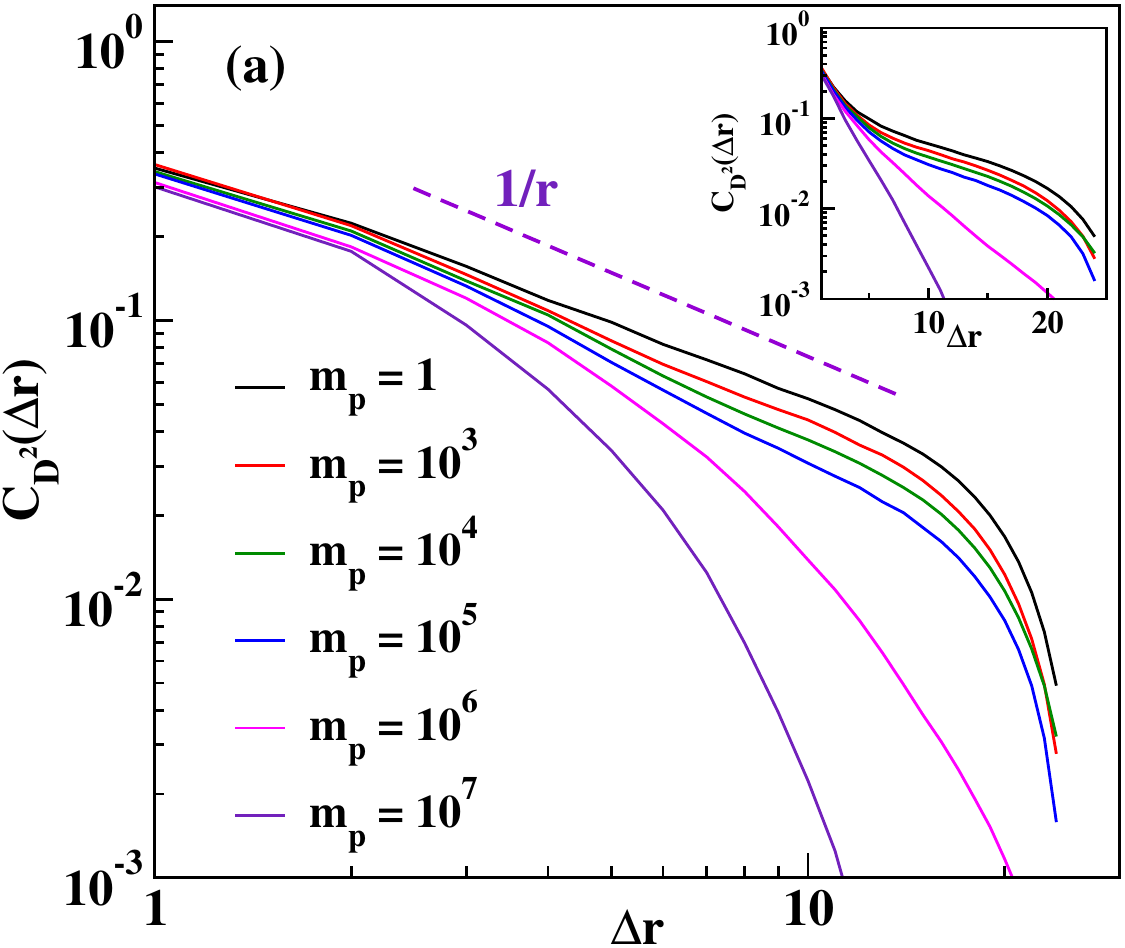}
	\includegraphics[width=60mm]{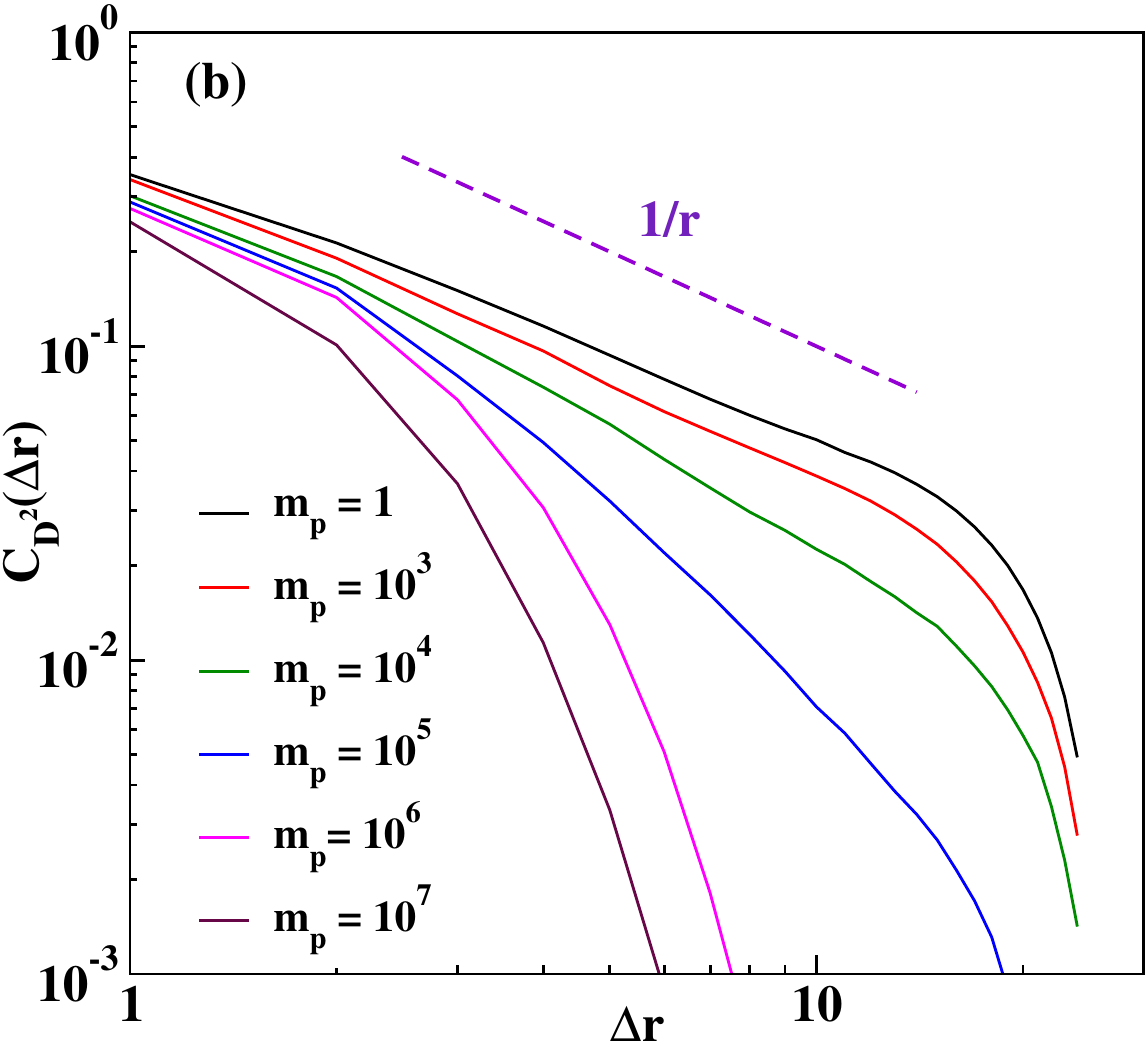}
	\includegraphics[width=60mm]{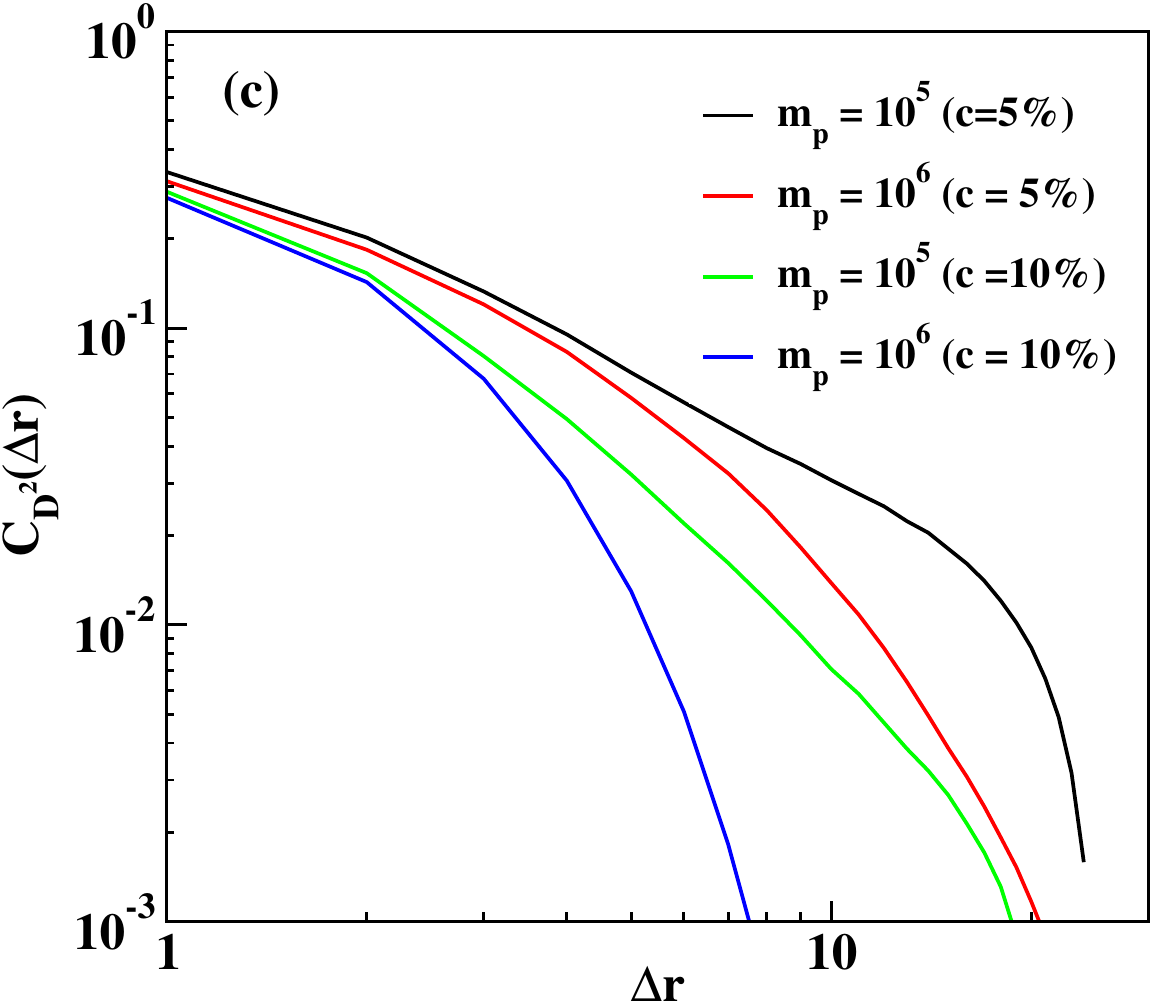}
	\caption{Spatial correlation function of the non-affine displacement field for different $m_P$ values for c = 5\% and 10\% are shown in (a) and (b) respectively in the log-log scale. Inset of (a) represents the results on the semi-log scale (see text). In (c) we show a comparison of the results in (a) and (b) for $m_p = 10^5$ and $10^6$.}
	\label{d2}
\end{figure}

\indent To understand the role of shear rate on the threshold mass, we repeat the same exercise mentioned above for two more different choice of $\dot{\gamma}=10^{-5}$ and $10^{-3}$. The corresponding spatial distributions of the $D^2$ field are shown in Fig.~\ref{snaperate}. The results indicate the critical mass to be $m_c  = 10^6$ and $10^4$ respectively. Therefore, the critical mass decreases with increasing shear rate.  \\
\indent To quantify the geometry of plastic events over space, we compute the spatial correlations of the $D^2$ field as follows:
\begin{equation}
	C_{D^2}(\Delta \textbf{r}) = \frac{\langle D^2(\textbf{r}+\Delta \textbf{r})  D^2(\textbf{r})\rangle-\langle D^2(\textbf{r})\rangle^2}{\langle D^2(\textbf{r})^2\rangle-\langle D^2(\textbf{r})\rangle^2}
	\label{cor-d2}
\end{equation}
Here the angular brackets represent an average of spatial coordinate $\textbf{r}$ over 50 independent samples. The result is shown in Fig.~\ref{d2} for the pure system in the steady state regime.
As expected, the $C_{D^2}(\Delta \textbf{r})$ shows a long-range correlation exhibiting power law behavior of $1/r$ in the plastic regime \cite{pinaki2023}. However, in the presence of impurity, increasing the $m_P$ gradually reduces the extend of correlation range of the non-affine displacement field. Beyond the critical mass $m_c$, the correlation completely deviates from long-range spatial order, and transforms from the power-law to an exponential decay. The exponential nature is demonstrated in the inset of Fig.~\ref{d2}a, where the data set appears to be linear on the semi-log scale. This provides the evidences for the crossover from a system-spanning shear band to localized plastic activities throughout the system \cite{pinaki2023,meenakshi,nikolai2016,nikolai2017}. This observation confirms the complete localization of the plastic events. Increase in impurity concentration for a chosen $m_P$ also shows a similar change in the correlation function. However, the critical mass for the transition is significantly lower for the higher concentration. This is demonstrated in Fig.~\ref{d2}c. For the 5\% disorder concentration, the transition from a scale-free power law behavior to a completely exponential decay occurs at a $m_c \approx 10^5$. In contrast, with a 10\% concentration, this transition happens much earlier, at a $m_c \approx 10^4$. This confirms that a higher disorder concentration leads to a drop in the critical mass required for such transition. 
\subsection{Strain Fluctuations}
\begin{figure}[hbt]
	\centering
	\includegraphics[width=42.5mm,height=37mm]{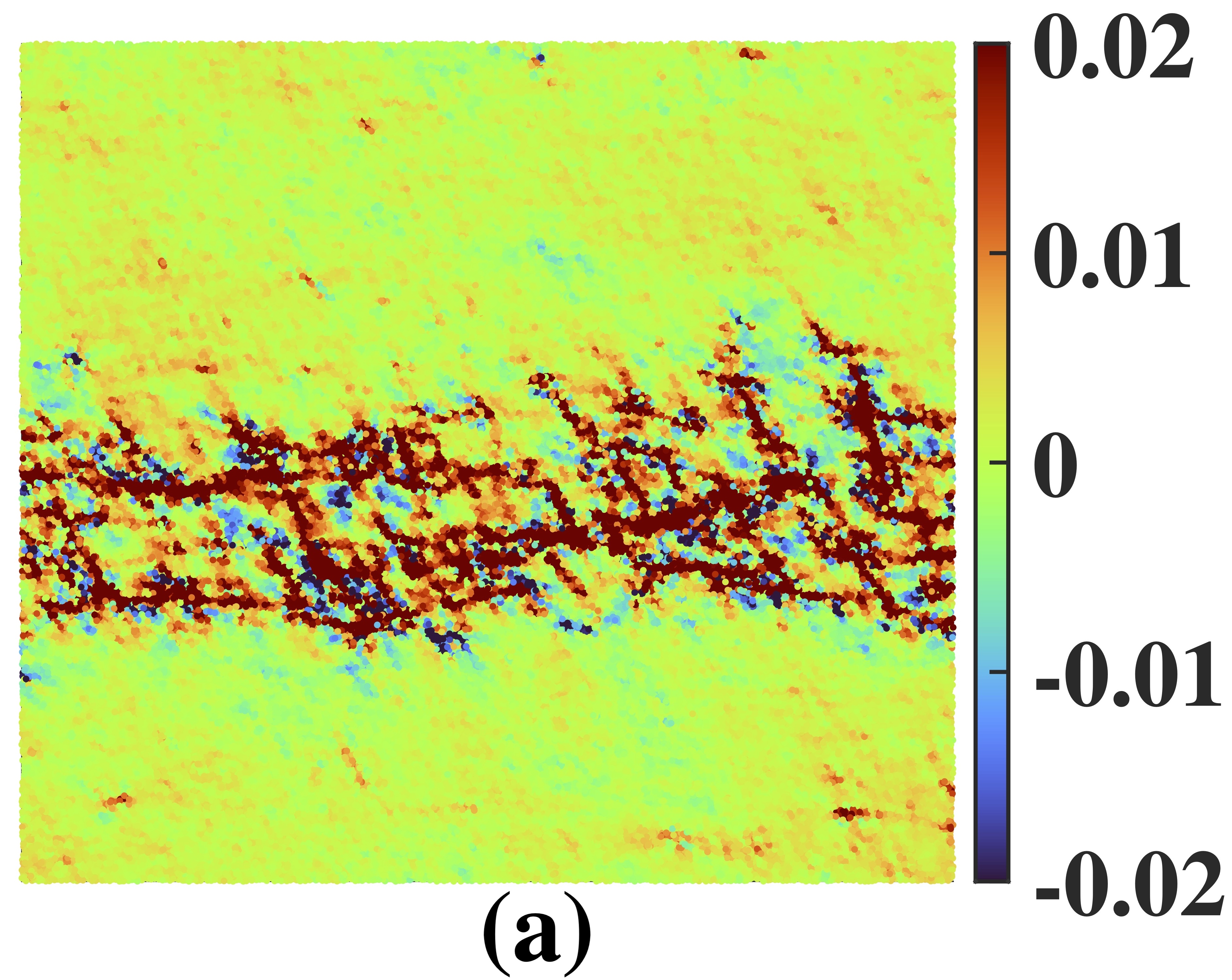}
	\includegraphics[width=42.5mm,height=37mm]{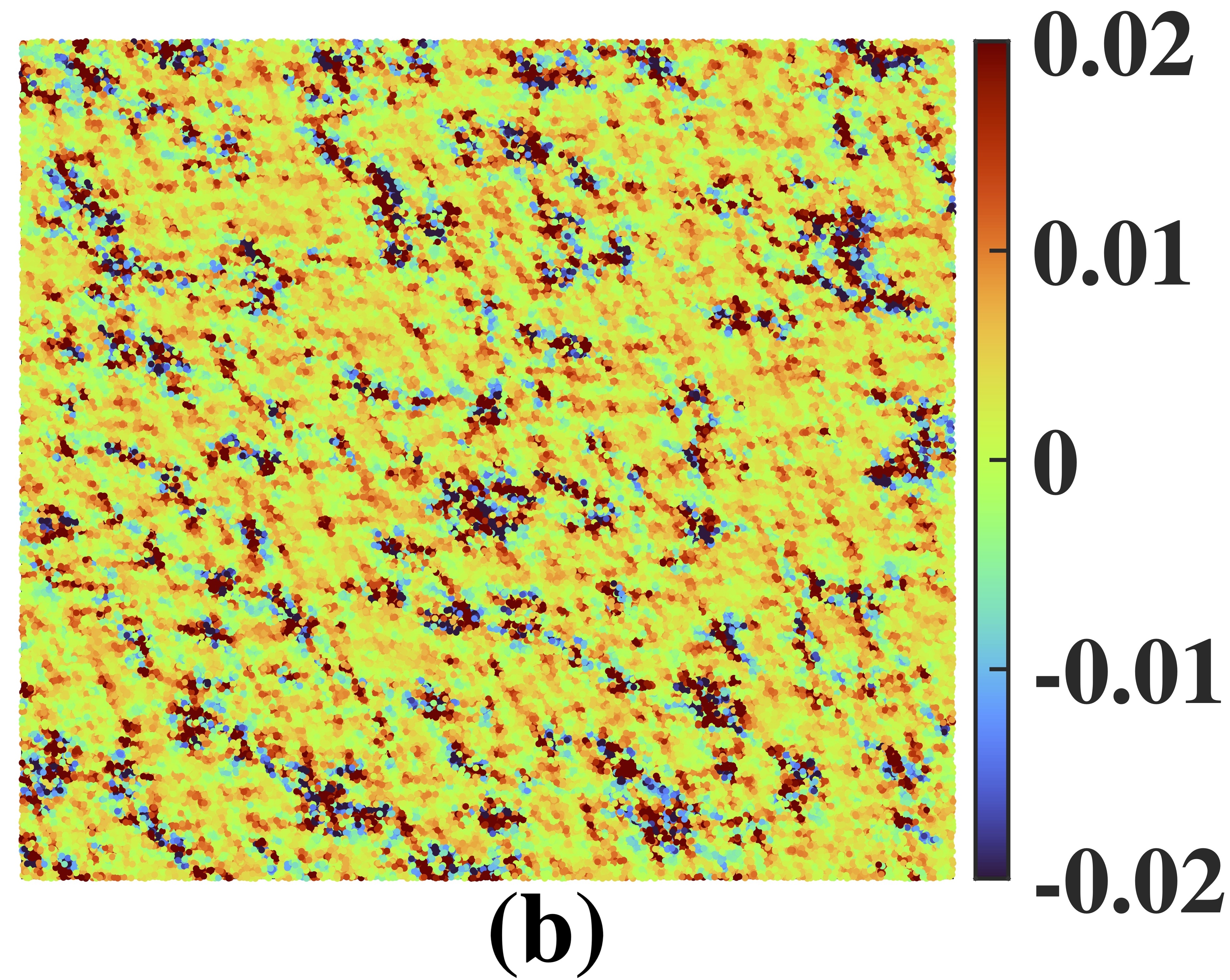}
	\caption{ The spatial distribution of the shear strain in the steady state regime for $m_p$ = 1 and $10^6$ are shown in (a) and (b) respectively.}
	\label{straindistribution}
\end{figure}

It is demonstrated in several experiments and simulations that the plastic deformation in amorphous solids occurs due to localized rearrangement of particles associated with long-range quadrupolar strain field \cite{chikkadi2011,Chattoraj,Chikkadi2015,Hassani}. The strain field resembles the Eshelby elastic field around an inclusion in a homogeneous isotropic solid \cite{eshelby}. The overall deformation arises from spatio-temporal interactions of such plastic rearrangements, mediated by elasticity. For further understanding the effect of annealed disorder on the underlying spatial organization of the deformation fields in amorphous solids, in this section we focus on the local strain tensor $\epsilon_{ij}$ which corresponds to the symmetric part of $\boldsymbol{\mathrm{J}}$ in Eq.~\ref{eq-d2}. Therefore, we compute the principal shear strain component $\epsilon_{xz}$ for our deformed systems. Fig.~\ref{straindistribution} illustrates the distribution of $\epsilon_{xz}$ in the plastic regime for a pure system and a 5\% disordered system with $m_p=10^6$ ($>m_c$).
 Particles are color-coded according to their strain values, ranging from negative to positive. Positive strain values indicate locations of high strain activity zones, which undergo large, irreversible rearrangements known as shear transformation zones, surrounded by an elastic matrix. In Fig.~\ref{straindistribution}a we observe an inhomogeneous shear strain field in the pure system. On the contrary, increasing $m_P$ of the impurity particles gradually suppress the shear band formation and promote a more homogeneous strain distribution. Eventually, beyond a critical mass of $m_p \approx 10^5$, the accumulated strain distribution separates into individual, independent shear transformation zones in a homogeneous elastic solid as shown in Fig.~\ref{straindistribution}b.
 
\begin{figure}[ht]
	\centering
	\includegraphics[width=70mm]{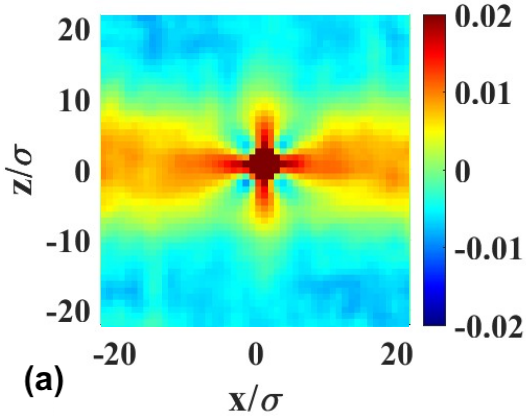}\\
	\includegraphics[width=40mm]{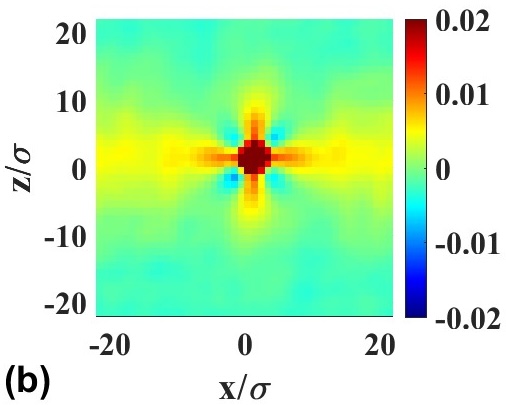}
	\includegraphics[width=40mm]{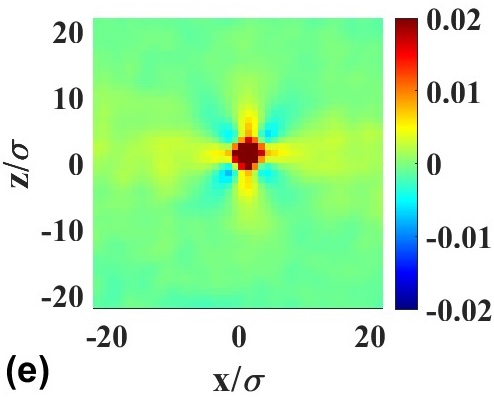}
	\includegraphics[width=40mm]{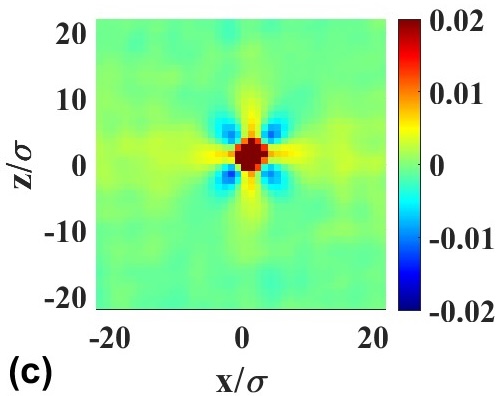}
	\includegraphics[width=40mm]{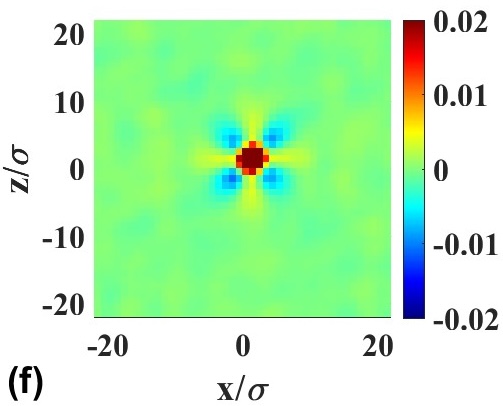}	\includegraphics[width=40mm]{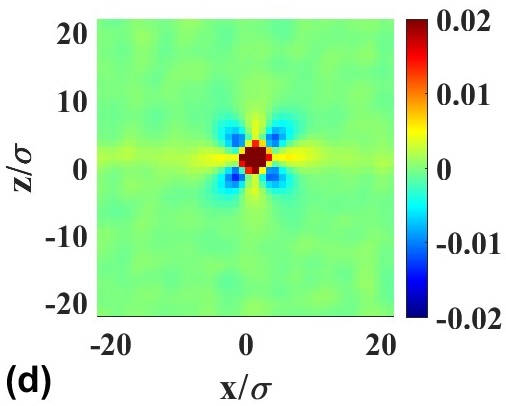}	\includegraphics[width=40mm]{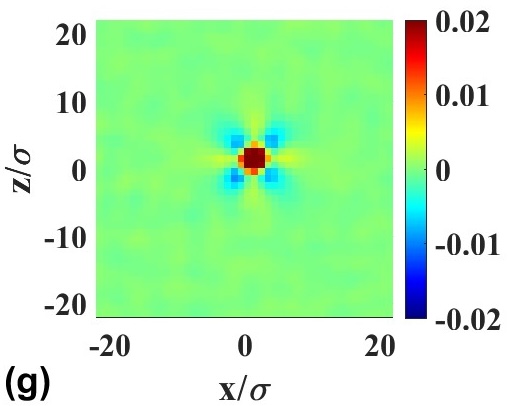}
	\caption{Spatial strain correlation function in the plastic regime for (a) the pure system, (b)-(d) c=5\%, $m_P = 10^4, 10^5$ and $10^6$, (e)-(g) c = 10\%, $m_P = 10^4, 10^5$ and $10^6$ respectively.}
	\label{quadrupolar}
\end{figure}

Next, we investigate the strain fluctuation by computing  the spatial correlation of the principal shear strain component $\epsilon_{xz}$ given as 
\begin{equation}
	C_{\epsilon}(\Delta \textbf{r}) = \frac{\langle \epsilon_{xz}(\textbf{r}+\Delta \textbf{r}) \epsilon_{xz}(\textbf{r})\rangle-\langle \epsilon_{xz}(\textbf{r})\rangle^2}{\langle \epsilon_{xz}(\textbf{r})^2\rangle-\langle \epsilon_{xz}(\textbf{r})\rangle^2}
	\label{cor-eps}
\end{equation}
For the pure system, the strain correlation shows distorted quadrupolar symmetry in the plastic regime, as shown in Fig.~\ref{quadrupolar}. 
This is due to the formation of sharp shear band spanning the entire system. On increasing the mass of the impurity particles, the dissociation from the shear band formation initiates, resulting in the localized plastic events with the prominent emergence of a quadrupolar symmetry. For the impurity concentration $c=5\%$ and $10\%$ and the corresponding $m_p = 10^6$ and $10^5$ respectively, the transition from a distorted to a sharp quadrupolar symmetry is observed as shown in Fig.~\ref{quadrupolar}. On further increasing the mass, a perfect quadrupolar symmetry persists with the shear transformation zones (inclusions) isolated by the elastic activity, resembling the Eshelby inclusions in a homogeneous solid \cite{eshelby}.

\begin{figure}[hbt!]
	\centering
	\includegraphics[width=75mm,height=60mm]{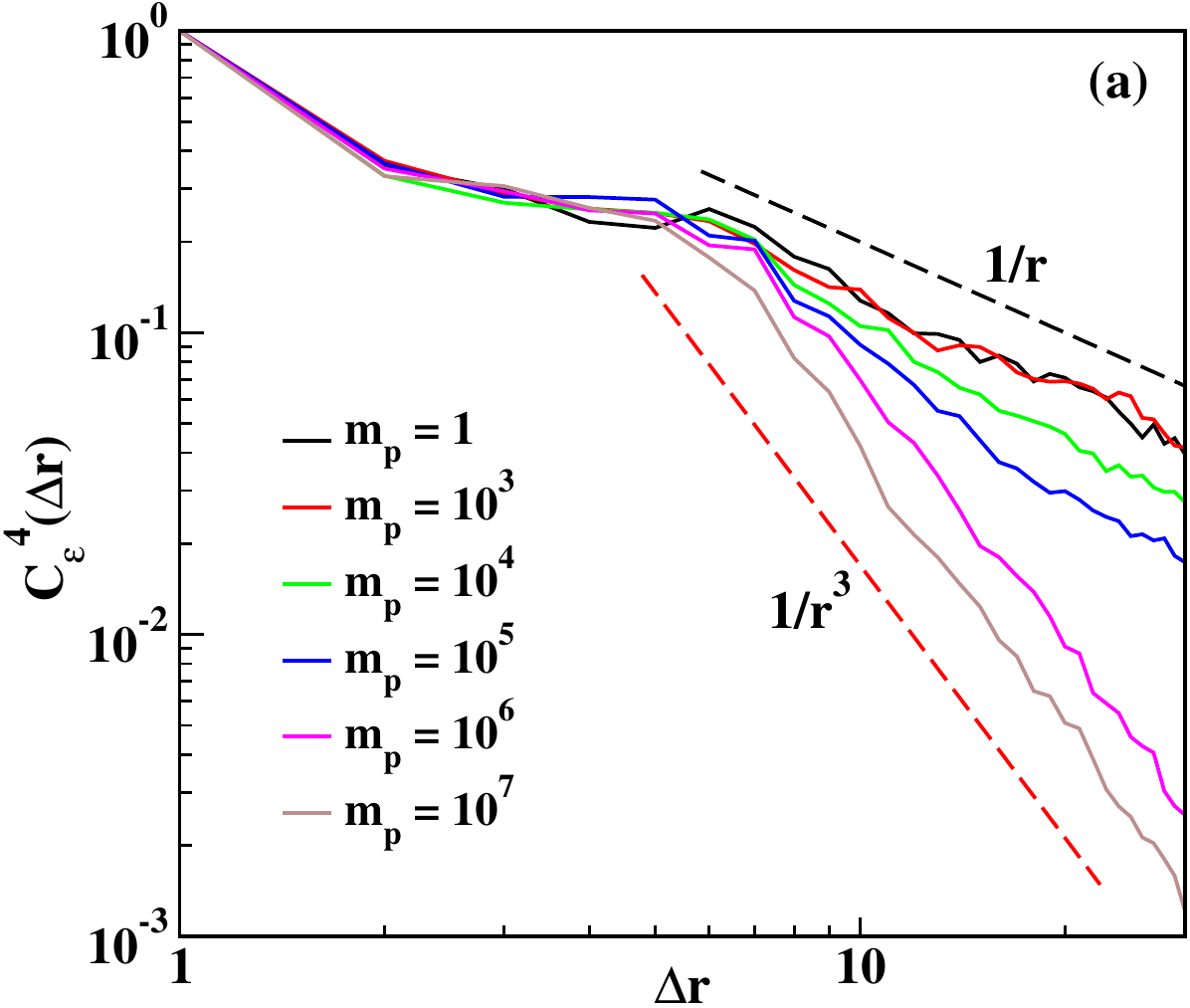}
	\includegraphics[width=75mm,height=60mm]{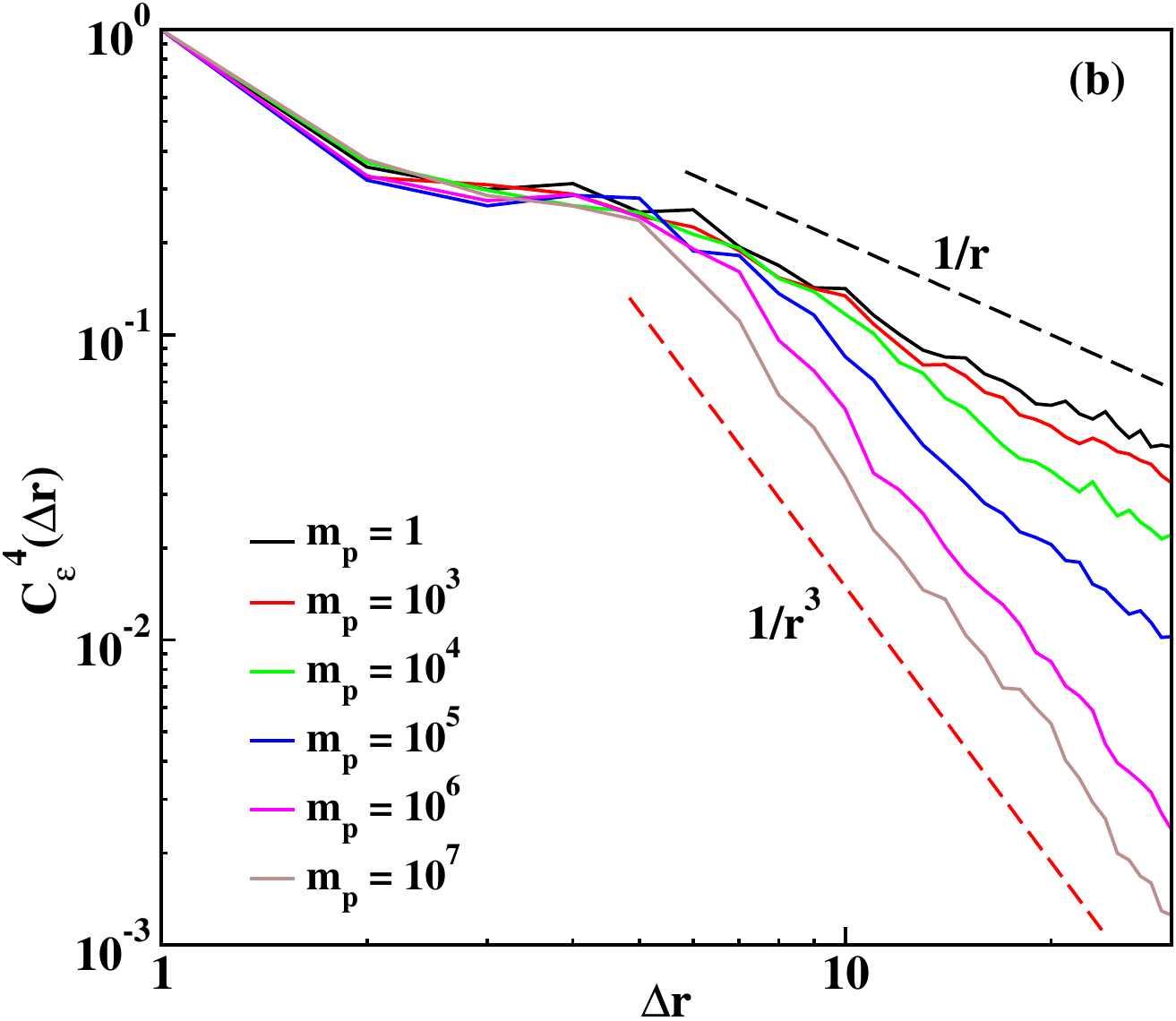}
	\caption{The radial decay of the strain correlation projected on to the circular harmonics for the system with two different impurity concentration c= 5\% and 10\% for different $m_p$ values are shown in (a) and (b) respectively. The dashed lines represent the guideline for the slope.}
	\label{straincorrelation}
\end{figure}

\begin{figure}[hbt!]
	\centering
	\includegraphics[width=75mm]{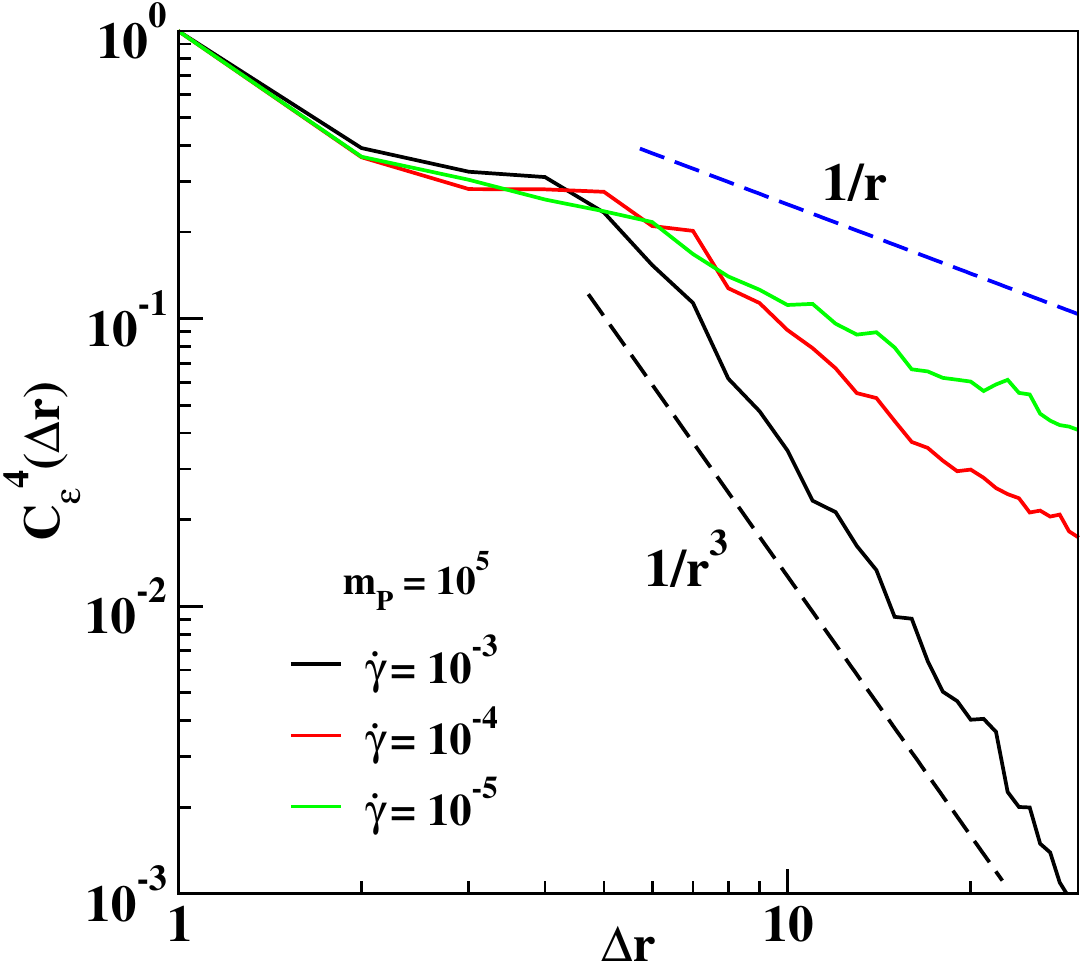}
	\caption{The comparison of the radial decay of the strain correlation projected on to the circular harmonics for our system with c= 5\% and $m_p = 10^5$ for three different strain rate $\dot{\gamma} = 10^{-3}, 10^{-4}$ and $10^{-5}$. The dashed lines represent the guideline for the slope.}
	\label{compare-rate}
\end{figure}

Next, we compute the radial correlation of the strain by projecting it on to the corresponding circular harmonic given by
\begin{equation}
	C^{\epsilon}_4(\delta r) = \int_0^{2\pi} C^{\epsilon}_4(\delta r,\theta) \cos{(4\theta)} d\theta
	\label{cor-sph}
\end{equation}
Fig.~\ref{straincorrelation} shows the radial dependence of strain correlation following a power-law decay of $1/r$ for the pure system in the plastic regime. This decay indicates a long-range elastic strain field which is slower than the $1/r^3$ behavior (shown as a dashed line in the same figure) observed for Eshelby inclusions for a single strained spherical inclusion in isotropic elastic solid \cite{Hassani, Sagar}. The $1/r$ decay originates from the strain correlations that capture the elastoplastic response of not a single inclusion but several interacting inclusions \cite{Chikkadi2015, Sagar}. For the system with annealed disorder, for smaller $m_P$ the decay of $C^{\epsilon}_4(\delta r)$ in the plastic regime is similar to that of the pure system. But, as the $m_P$ is increased, the $C^{\epsilon}_4(\delta r)$ starts to deviate from the $1/r$ behavior and becomes shorter ranged. The change occurs faster in the systems with higher impurity concentration. Eventually, above the critical mass of the impurity particles, the $1/r^3$ decay of the $C^{\epsilon}_4(\delta r)$ is observed in the plastic regime which resembles the behavior of the pure system in the elastic regime. This type of transition from a slower ($1/r$) to a faster ($1/r^3$) decay observed in our study with increasing annealed disorder particles mass is attributed to the gradual localization of plastic events, as shown in Figs. \ref{straincorrelation}(a) and (b) for $c=5\%$ and 10\%, respectively. The critical mass above which the decay follows $1/r^3$ behavior are $m_c = 10^5$ and $10^4$ for $c=5\%$ and 10\%, respectively. These results are consistent with the predictions of the $m_c$ values from the $D^2$ correlation functions. \\   
\indent	Finally, we revisit the effect of shear rate on the plastic events of the disordered system in terms of the strain correlation. Therefore, we compute the $C^{\epsilon}_4(\delta r)$ for our glassy system with c = 5\% in the plastic regime deformed with three different strain rate $\dot{\gamma} = 10^{-3}, 10^{-4}$ and  $10^{-5}$. The results are shown in Fig.~\ref{compare-rate} for a fixed $m_P=10^5$. For a low strain rate of $\dot{\gamma} = 10^{-5}$ a long range decay of $1/r$ is observed, indicating the existence of shear band. On the other hand, when the sample is deformed at a faster rate ($\dot{\gamma} = 10^{-3}$), the decay exhibits $1/r^3$ nature. This corresponds to the localized plastic events and complete suppression of shear band. For $\dot{\gamma} = 10^{-4}$ an intermediate decay nature is observed. These results are consistent with our observations in Fig.~\ref{snaperate}. Therefore, we can convincingly conclude that the critical mass decreases with increasing the rate of deformation. \\
\section{Conclusions}
In conclusions, we have studied the effect of annealed disorder on the mechanical properties and plasticity of a modeled amorphous solid. The nature of disorder considered here closely resembles the real experimental systems and does not alter the total degrees of freedom and the potential energy landscape of the original system. The disorder particles are ergodic and take part in the non-affine displacement during external loading. Two different impurity concentrations are considered. The mechanical properties of such system are explored at the microscopic level. A significant enhancement of the toughness, shear modulus and the yield stress is observed with increasing impurity mass and well as the concentration. The study of the spatial correlation of the non-affine displacement field demonstrated that the system spanning shear band formation was gradually suppressed with increasing mass of the annealed disorder particles. We identified a critical mass above which the plastic events become completely localized in the plastic regime. The value of the critical mass decreases with the impurity concentration. For better understanding of the above mentioned results we resorted to the correlations of local strain fluctuations. Starting from a distorted quadrupolar structure of the strain correlation in the plastic regime of the pure system, a perfect quadrupolar symmetry was recovered with annealed disorder. This was accompanied with the transition of the strain correlation from the long ranged $1/r$ decay to a short ranged $1/r^3$ behavior resembling the elastic regime of the pure system. Finally, the effect of shear rate on the shear band formation is explored in the presence of disorder. The critical mass of the impurity particles is found the decrease with increasing the rate of deformation. Our model for amorphous materials featuring annealed disorder empowers us with enhanced control over adjusting material strength, leveraging the finite mass of impurities. This capability opens avenues for designing novel materials tailored to specific properties.

\section*{Acknowledgements} 
B. Sen Gupta acknowledges Science and Engineering Research Board (SERB), Department of Science and Technology (DST), Government of India (no. CRG/2022/009343) for financial support. Meenakshi L. acknowledges VIT for doctoral fellowship. We gratefully acknowledge Vijayakumar Chikkadi for many useful discussions and providing us the code to calculate strain correlations.

\end{document}